\pdfoutput=1
\documentclass[fleqn,10pt]{wlscirep}
\usepackage[utf8]{inputenc}
\usepackage[T1]{fontenc}
\usepackage{booktabs}
\usepackage{mathtools}
\usepackage{comment}
\usepackage{dcolumn}
\usepackage{multirow}
\usepackage{float}
\PassOptionsToPackage{hyphens}{url}\usepackage{hyperref}
\usepackage{subfiles}
\usepackage{parskip}
\usepackage{graphicx}
\usepackage{amssymb}
\usepackage{url}
\usepackage{subfigure}
\usepackage{lineno}
\usepackage{color}
\usepackage{colortbl}
\usepackage{fancyhdr}
\usepackage{ctable}
\definecolor{Gray}{gray}{0.9}
\usepackage{enumitem}
\usepackage{caption}
\usepackage{array}
\usepackage{makecell}
\usepackage{cellspace}
\usepackage{tabularx}

\newcommand{\comments}[1]{\textcolor{cyan}{[(comment) #1]}}

\title{Insider stories: Analyzing Internal Sustainability Efforts of major US companies from online reviews}

\author[1]{Indira Sen}
\author[2,3*]{Daniele Quercia}
\author[4]{Licia Capra}
\author[5]{Matteo Montecchi}
\author[2]{Sanja Šćepanović}

\affil[1]{GESIS-Leibniz Institute for Social Sciences, Germany}
\affil[2]{Nokia Bell Labs, Cambridge, UK}
\affil[3]{CUSP, King's College, London, UK}
\affil[4]{University College London, UK}
\affil[5]{King's College Business School, London, UK}
\affil[*]{quercia@cantab.net}

\begin{abstract}
It is hard to establish whether a company supports internal sustainability efforts (ISEs) like gender equality, diversity, and general staff welfare, not least because of lack of methodologies operationalizing these internal sustainability practices, and of data honestly documenting such efforts. We developed and validated a six-dimension framework reflecting Internal Sustainability Efforts (ISEs), gathered more than 350K employee reviews of 104 major companies across the whole US for the (2008-2020) years, and developed a deep-learning framework scoring these reviews in terms of the six ISEs.  
Commitment to ISEs manifested itself at micro-level --- companies scoring high in ISEs enjoyed high stock growth. This new conceptualization of ISEs offers both theoretical implications for the literature in corporate sustainability, and practical implications for companies and policy makers. To further explore these implications, researchers need to add potentially missing ISEs, to do so for more companies, and establish the causal relationship between company success and ISEs.

\end{abstract}

\begin{document}

\flushbottom
\maketitle
\thispagestyle{empty}

\section*{Introduction}

\begin{figure}[t]
\centering
\includegraphics[width=0.5\linewidth]{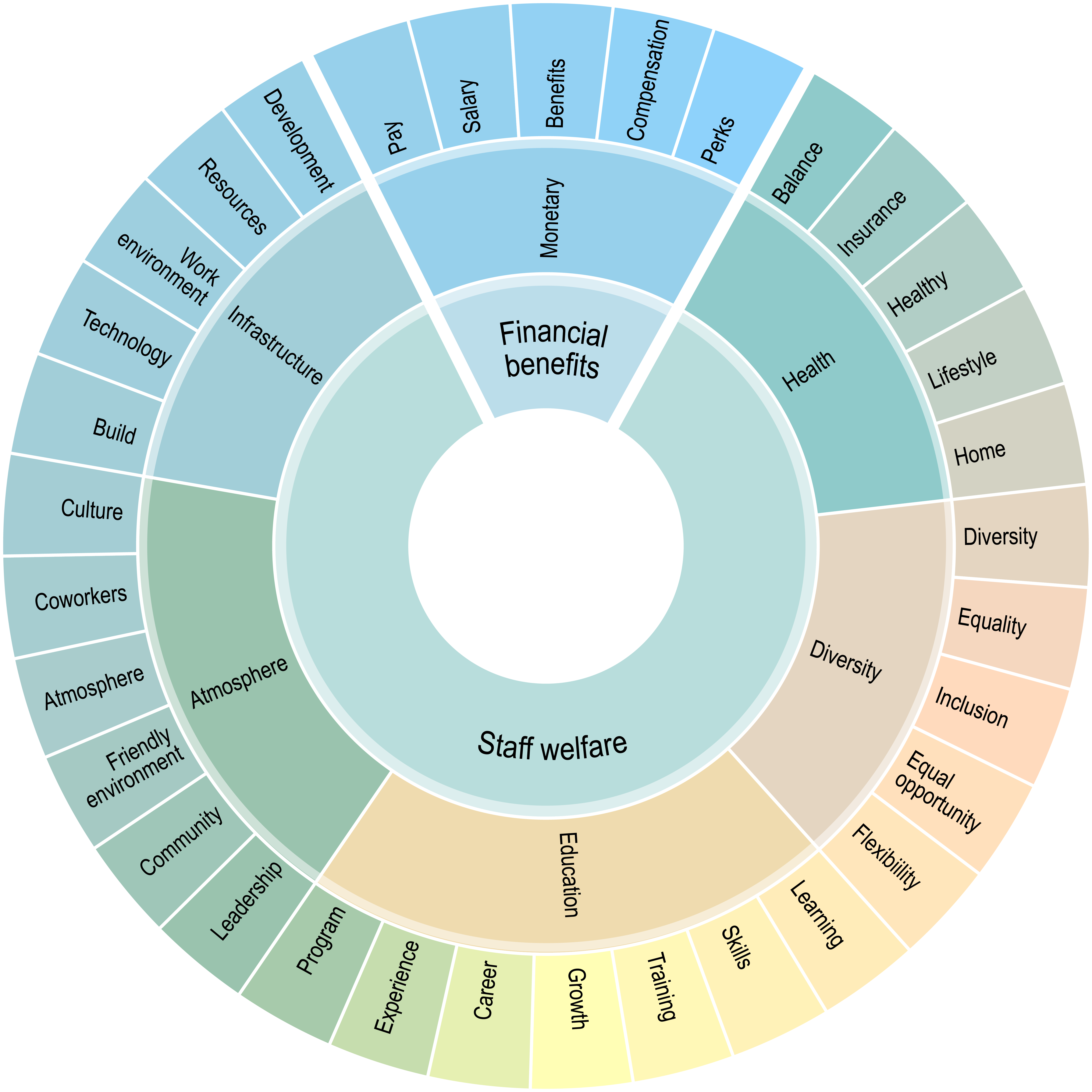}
\caption{\textbf{The wheel of Internal Sustainability Efforts (ISEs).} The wheel includes the two 
macro categories (financial benefits \emph{vs.} staff welfare) under which  the six ISEs are classified. The wheel's outer layer reports 
keywords representative of each ISE.}
\label{fig:wheel}
\end{figure}

Investments in sustainability are becoming paramount as many companies are under constant pressure to reduce the social and environmental impact of their operations~\cite{bai2020supply} and increase accountability towards stakeholders and the wider society ~\cite{de2021reimagining,serafeim2020social,wang2022social}. While corporate sustainability efforts tend to focus primarily on external stakeholders (e.g., customers, supply-chain partners, governmental organizations)~\cite{doi:10.1177/0022242921992052}, internal stakeholders (e.g., employees) represent a critical, and sometimes overlooked, target group to ensure effective corporate engagement with the sustainability agenda~\cite{martin2021exploring,paine2014sustainability,chatzopoulou2022corporate}. 

Internal sustainability efforts (ISEs) encompass a wide range of corporate policies directed towards internal stakeholders, including, for example, promoting a healthy employee work-life balance~\cite{kelliher2019all}, investing in gender equality and diversity~\cite{NADEEM2017874}, and ensuring an harassment-free working environment~\cite{cassino2019race}. These ISEs can reduce staff turnover~\cite{giauque2019stress} and improve market competitiveness~\cite{wang2012explaining}. However, although many companies openly advertise their commitment to internal sustainability, employees often report contrasting accounts of their experience of such efforts~\cite{peloza2011how}, and the extent to which ISEs successfully propagate throughout the organization remains unclear.

We partly tackle those issues by running a large-scale assessment of organizational practices aligned with ISEs. Since there exists no agreed-upon definition of corporate ISEs, to inform our research we started from the United Nations (UN) World Commission on Environment and Development (WCED) definition of sustainability as a strategy oriented towards ``meeting the needs of the present without compromising the ability of future generations to meet their own needs''~\cite{wced1987world}. This definition is operationalized in the UN 17 Sustainable Development Goals (SDGs), which represent both a framework and a call-to-action for organizations to invest in addressing critical societal issues such as ``good health and well-being'', ``decent work and economic growth'', and ``peace, justice and strong institutions''~\cite{nations2015transforming}. Not all 17 UN SDGs are relevant to a company's {\em internal} stakeholders (this is the case, for example, for SDG ``life under water''). To identify the relevant ones and sharpen their definitions in the internal corporate context, we  developed and validated a mixed-method approach that ended up paraphrasing the broad UN SDGs into six corporate-relevant ISEs. These ISEs concerned  health, education, diversity, monetary benefits, infrastructure, and atmosphere (Figure~\ref{fig:wheel}). Core to the approach is a state-of-the-art Natural Language Processing (NLP) framework that processed more than 350K geo-referenced reviews about 104 S\&P 500 companies.

\section*{Data}
Our aim was to understand and capture the microfoundations of ISEs; we did so in a bottom-up fashion, starting from the perspectives of employees. More specifically, we collected data from a popular company reviewing site, where current and, more likely, former employees write reviews about their own corporate experiences, ranging from job interviews to salaries to workplace culture. These reviews have been recently used in studies exploring corporate culture at scale~\cite{das2020modeling}. As of 2021, there are 50M monthly visitors on the platform, and 70M reviews of 1.3M companies. To ensure quality reviews, the site: a) performs both automatic and manual content moderation; b) allows for full access to content only to users who register on the site and write at least one review each (encouraging neutral and unbiased reviews); and c) allows for posting maximum one review per employee per year. Our dataset consisted of reviews published over twelve years, from 2008 to 2020. 

Each review consists of a title; a `pro' portion (i.e., positive aspects of the company); a `con' portion (i.e., its negative aspects); a set of four ratings on a [0,5] scale scoring the company's \emph{balance}, \emph{career}, \emph{culture}, and \emph{management}; and a final \emph{overall} rating of the company. Since reviewers have the option to include their location, we were able to identify the states for part of the reviews. To ensure the robustness of our text processing method,  we retained companies that had at least 1,000 reviews and were present in at least 10 states, leaving us with a dataset of 358,527 reviews of 104 US-based companies (which represented 88.7\% of the original dataset); 80\% of these are S\&P 500. As detailed in Supplementary Information, these 104 companies offer the same level of representativeness as the S\&P 500 companies, in terms of the distribution of industry sectors and the geographic distribution across states. In addition to the reviews, we collected yearly stock growth values of the 104 companies from the Yahoo Finance Portal\cite{WinNT}.

\section*{Methods}

\subsection*{The three-step mixed-method approach for defining ISEs}

We developed a mixed-method approach to operationalize ISEs. This approach unfolded in three main steps, which are detailed in Supplementary Information and summarized here as follows (Figure~\ref{fig:goal_selection}):

\begin{figure}[t]
\centering
\includegraphics[width=\linewidth]{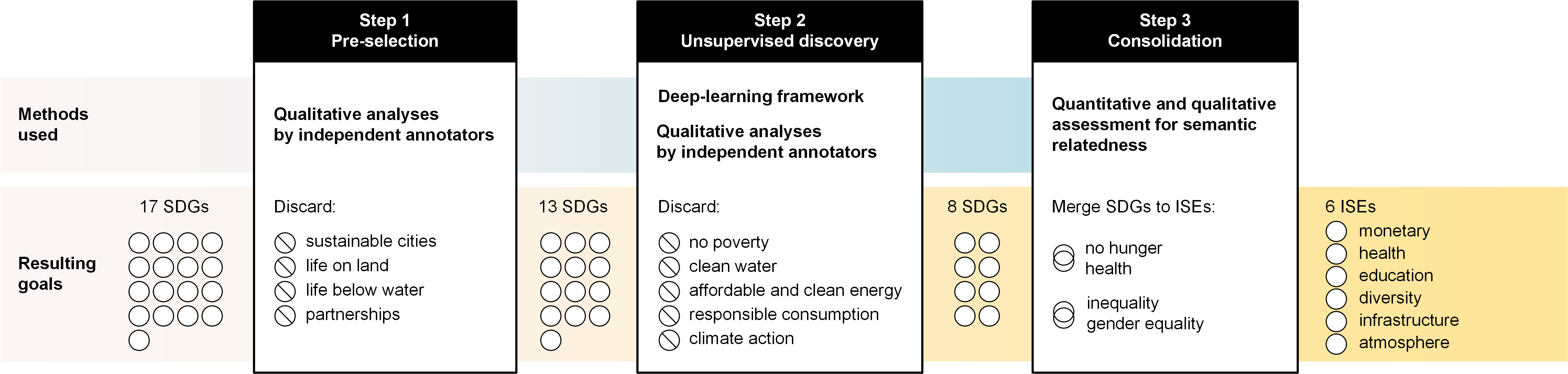}
\caption{\textbf{Summary of the three-step mixed-method approach for defining ISEs.} Starting with the 17 UN SDGs, three annotators unanimously discarded those that did not apply to the corporate context (step 1, pre-selection): 13 SDGs were left. From these, the subset of SDGs accurately captured by our NLP deep-learning framework  was identified (step 2, unsupervised discovery): 8 SDGs were selected.  Finally, three annotators merged the goals that, in the context of company reviews, ended up being paraphrased with very similar meanings (step 3, consolidation): this final step resulted in the identification of the six ISEs.}
\label{fig:goal_selection}
\end{figure}

\begin{description}
\item[Step 1 - Pre-selection of goals:] Using a deductive content analysis~\cite{elo2008qualitative}, three independent annotators assessed each of the UN seventeen goals' definitions and decided whether they applied to the corporate context or not. We took a conservative approach and discarded the goals that the annotators \emph{unanimously} discarded, which ended up  being four, leaving us with thirteen  potentially relevant goals (after step 1 in Figure~\ref{fig:goal_selection}). 

\item[Step 2 - Unsupervised discovery of  goals:] An unsupervised deep-learning framework based on the sentence-level BERT algorithm~\cite{reimers2019sentence} was developed (its technical  architecture is discussed in Supplementary Information). This framework scored each employee's review against the 13 goals found in the previous step. The framework identified the five  reviews most relevant to each goal, and three other independent annotators then manually assessed  the relevance of these reviews. To conservatively retain only the goals that were accurately identified by the framework,  we discarded any goal for which the majority of the annotators marked less than 4 of the goal's 5 reviews as relevant (overall, the agreement among the annotators was high, i.e., Fleiss $K=0.83$).  As a result, five goals were dropped; these had more to do with environmental sustainability (e.g., ``clean water'', ``climate change'') than with internal corporate affairs. This left us with eight goals (after step 2 in Figure~\ref{fig:goal_selection}).

\item[Step 3 - Consolidation of goals:] Finally, the three annotators assessed if any of the eight goals ended up acquiring very similar meanings in company reviews. Two pairs were merged, ultimately leaving us with six ISEs (after step 3 in Figure~\ref{fig:goal_selection}). Table~\ref{tab:goal_selection} reports the names of these ISEs (first column), corresponding original UN SDGs (second column), and related excerpts of real reviews (third column). 
\end{description}

\subsection*{Metrics}

We studied the six ISEs at the company-level $u$ to test whether commitment to ISEs manifests itself at a micro-level (e.g., in a company's growth). To that end, we computed the score $s(u, i)$ of the $i^{th}$ ISE for company $u$ as the  fraction of $u$'s reviews that mentioned $i$:

\begin{equation}
    s(u, i) = \frac{\sum_{p \in R(u)}sim_t(v_p,v_i)}{|R(u)|} 
\end{equation}

\noindent where $R(u)$ is the set of $u$'s reviews, $v_i$ is the SBERT (Sentence-BERT) vector of ISE $i$ (the six vectors/phrases for the ISEs are in Supplementary Information in Table 4), and $sim_t(v_p,v_i)$ is the \emph{thresholded} SBERT similarity score~\cite{reimers2019sentence} between the SBERT vector of review $p$ and the SBERT vector of ISE $i$. More precisely, $sim_t(v_p,v_i)$ is defined as:

\begin{equation}
    sim_t(v_p,v_i) = \begin{cases} sim(v_p,v_i), & \text{if } sim(v_p,v_i)>0.31 \textrm{ AND } sim(v_p,v_i) \textrm{>} 95\%(i)
    \\
    0,              & \text{otherwise}
    \end{cases}
\end{equation}

\noindent We chose the threshold of $0.31$ by computing the mean SBERT similarity for each of the 8 goals left after stage 2 of our 3-step ISE selection procedure as we had established that the NLP method worked well for these 8 goals. We then took the average value of the eight means (which was $0.31$). Based on further validation, we also established that the SBERT values for all ISEs were not equally distributed and, as such, the fixed generalized threshold of $0.31$ had to be paired with an ISE-specific threshold: based on our experiments reported in Supplementary Information, this latter threshold value (denoted as $95\%(i)$) was the 95\% percentile of the ISE's distribution, which is the very same threshold found in previous studies~\cite{choi20ten}. We finally ranked companies by their score $s(u,i)$ for each $i^{th}$ ISE. Note that, by review, we mean the pro portion of the review.  That is because we were mostly interested in positive initiatives (pros) rather than shortcomings (cons). In Supplementary Information, we indeed show that, if we were to instead take cons (or combine cons with pros together), our deep-learning framework would perform worse in the two validation steps of our mixed-method approach (steps 2 and 3).

\begin{table}[t]
\centering
\small
\begin{tabular}{@{}lll@{}}
\toprule
Internal Sustainability Efforts (ISEs)                                                                               & UN Goal                                                                             & Example of Review Sentence                                                                                                     \\ \midrule
\begin{tabular}[c]{@{}l@{}}Monetary \end{tabular}                                 & \begin{tabular}[c]{@{}l@{}}decent work and \\ economic growth\end{tabular}          & \begin{tabular}[c]{@{}l@{}}``Professional growth, training, co-workers, mutuality,\\  income, entrepreneurship''.\end{tabular}                \\
Health                                                                                          & good health and wellbeing                                                           & \begin{tabular}[c]{@{}l@{}}``Excellent work-life balance. Great information \\ offered to improve health and equality of life''.\end{tabular} \\
\begin{tabular}[c]{@{}l@{}}Education \end{tabular}                                & quality education                                                                   & \begin{tabular}[c]{@{}l@{}}``Encourage continual education and offer multiple \\ learning opportunities''..\end{tabular}                      \\
\begin{tabular}[c]{@{}l@{}}Diversity \end{tabular} & gender equality                                                                     & ``Respect for gender equality''.                                                                                                              \\
\begin{tabular}[c]{@{}l@{}}Infrastructure \end{tabular}                       & \begin{tabular}[c]{@{}l@{}}industry, innovation,\\  and infrastructure\end{tabular} & ``Good infrastructure to support the work environment''.                                                                                      \\
\begin{tabular}[c]{@{}l@{}}Atmosphere \end{tabular}                            & \begin{tabular}[c]{@{}l@{}}peace, justice, and \\ strong instituions\end{tabular}   & \begin{tabular}[c]{@{}l@{}}``Collaborative environment, excellent benefits, \\ opportunity for growth and development''.\end{tabular}         \\ \bottomrule
\end{tabular}
\caption{\textbf{The six internal sustainability efforts resulting from the three-step mixed-method approach for defining ISEs.}}
\label{tab:goal_selection}
\end{table}

\section*{Results}

\begin{figure}[t]
\centering
\includegraphics[width=0.7\linewidth]{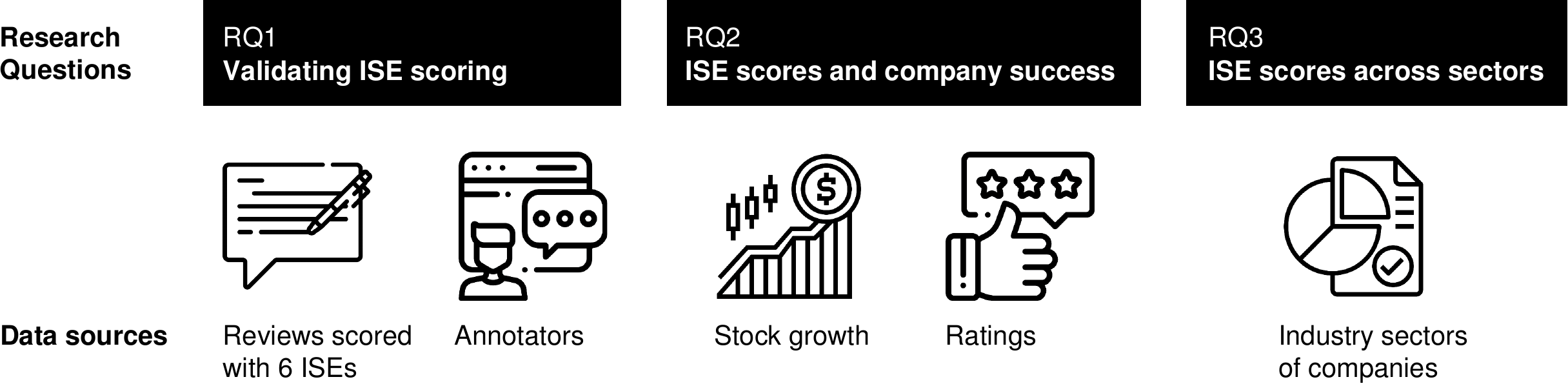}
\caption{Overview of the research questions investigated in this work and the sources of data we used for them.}
\label{fig:ise_steps}
\end{figure}

We identified each ISE's keywords from all reviews associated with it (e.g., the keyword `salary' for the `monetary' ISE), and  ascertained through a principled linguistic validation that the keywords are semantically related to the ISE (RQ1). After establishing that our ISE scoring is valid, we scored the companies, and studied the relationship between a company's ISE scores and its success in the forms of company ratings and stock growth (RQ2), and uncovered ISE scores variability across industry sectors (RQ3). Figure~\ref{fig:ise_steps} summarizes our analyses and the data used for them.

\subsection*{RQ1: Does our machine learning method capture Internal Sustainability Efforts?}\label{sec:validation}

We validated our deep learning method for detecting ISEs based on a triangulation approach~\cite{denzin2012triangulation}, during which we first established its face validity by inspecting the language used in reviews, and subsequently examined our results with respect to external reports. We discuss the former next, while the latter is detailed in Supplementary Information. 

To establish the face validity of the proposed ISE detection method, we took the linguistic approach explored by Das Swain et al.~\cite{das2020modeling} First, for each of the six ISEs, we obtained the most frequent keywords --- 1, 2, 3, and 4-grams from the reviews deemed relevant by our method. We then computed the TF-IDF scores for such n-grams, where each document was comprised of all shortlisted reviews for each ISE. Finally, we ranked keywords for each ISE based on their TF-IDF score. This allowed us to find the keywords judged to be important for a certain ISE by our embedding-based method. The top ranked keywords for the six ISEs are visualized as a heatmap in Figure~\ref{fig:tf_idf}. 

\begin{figure}[t!]
\centering
\caption{\textbf{Top n-grams in sentences expressing ISEs.} Darker colors (higher normalized TF-IDF score) indicate greater relative relevance to a particular ISE.}
\label{fig:tf_idf}
\end{figure}

We observed many keywords to be highly discriminative of the ISE they associated with: for example, keywords `pay good' and `salary' were (correctly) ranked highly for ISE `monetary' only; `health', `health benefits', and `take care' were ranked highly for ISE `health' instead. Keywords `opportunities learn', `experience', `good train', and `program' were uniquely strongly associated with ISE `education'. Keyword `flexibility' was highly discriminative of ISE `diversity'; `industry' and `technology' were strongly associated with ISE `infrastructure'; lastly, n-grams like `positive work environment' and `friendly work environment' strongly associated with ISE `atmosphere'. Other keywords ranked highly in more than one ISE instead: this was the case, for example, for keywords `benefit worklife balance' and `family', which were highly associated with both the `health' ISE (as one might expect), and to the `diversity' ISE. Health-enhancing factors like work-life balance and flexible working conditions options have been shown to facilitate gender equality and improve the diversity of employees~\cite{chung2018flexible,lyonette2015part}, therefore it was promising that our method was capable of picking up these  semantically-related concepts too.

Indeed the 6 ISEs we identified were not mutually exclusive concerns (and neither are the UN SDGs), and one may wonder to what extent they are semantically related. To shed light onto this question, we conducted a Principal Component Analysis (PCA) on \textrm{s(u, i)} at a company level to assess how much of the variance in the data could be explained by different principal components, and how those components related to the 6 ISEs. We found that, at company level, just two components explained 88\% of the variance --- specifically, the first component explained 73\% and the second component explained 15\%. We report the correlation between the first two PCA components and the six ISEs in the last two columns of Table~\ref{tab:pca}. 

We observed that all ISEs with the exception of `monetary' were strongly correlated with the first component and weakly negatively correlated with the second component; on the other hand, `monetary' was moderately correlated with both the first and second principal components. These two findings suggested that the `monetary' ISE was orthogonal to the other five, and that these other five were strongly interconnected with one another. Indeed, one may expect that improving work-life balance has a positive impact on both the `health' ISE \textit{and} the `diversity' ISE; on the other hand, improving monetary conditions may not directly affect other aspects of corporate internal sustainability. Overall, we thus found two main facets of employee-centred sustainability --- a \emph{staff welfare}-related one ($PC_1$) and a \emph{financial benefits}-related one ($PC_2$). To avoid multicollinearity, we used these two main facets of ISEs (rather than the six individual ones) to answer the following research questions.

\begin{table}
\centering
\scriptsize
\begin{tabular}{llccccccc}
\hline
                         & Monetary & Health & Education & Diversity & Infrastructure & Atmosphere & \begin{tabular}[c]{@{}c@{}}Staff Welfare \\ (PC1)\end{tabular} & \begin{tabular}[c]{@{}c@{}}Financial Benefits\\ (PC2)\end{tabular} \\ \hline
Monetary                 & 1.00     & 0.74   & 0.39      & 0.43      & 0.41           & 0.55       & \textbf{0.67}                                                  & \textbf{0.68}                                                      \\
Health                   & 0.74     & 1.00   & 0.55      & 0.76      & 0.74           & 0.85       & 0.90                                                           & 0.32                                                               \\
Education                & 0.39     & 0.55   & 1.00      & 0.72      & 0.77           & 0.73       & 0.82                                                           & -0.39                                                              \\
Diversity                & 0.43     & 0.76   & 0.72      & 1.00      & 0.82           & 0.92       & 0.90                                                           & -0.19                                                              \\
Infrastructure           & 0.41     & 0.74   & 0.77      & 0.82      & 1.00           & 0.92       & 0.90                                                           & -0.25                                                              \\
Atmosphere               & 0.55     & 0.85   & 0.73      & 0.92      & 0.92           & 1.00       & 0.96                                                           & -0.07                                                              \\
Staff Welfare (PC1)      & 0.67     & 0.90   & 0.82      & 0.90      & 0.90           & 0.96       & 1.00                                                           & 0.00                                                               \\
Financial Benefits (PC2) & 0.68     & 0.32   & -0.39     & -0.19     & -0.25          & -0.07      & 0.00                                                           & 1.00                                                               \\ \hline
\end{tabular}
\caption{Cross-correlation between the 6 ISE scores and the 2 principle components obtained via PCA at a company level.}\label{tab:pca}
\end{table}

\subsection*{RQ2: Is sustainability associated with company success?} 
\noindent
There are several ways to measure a company's success. We considered two complementary ones: the online ratings it received from its employees (available from the company reviewing site), and its financial position (measured as stock growth).\\

\noindent
\textit{Sustainability and company online ratings.} Employees have the option to rate the company they are reviewing based on four different facets --- balance, career, culture, management, plus a fifth company's overall one. We thus investigated to what extent a company's success across these five facets could be predicted based on the company's commitments to the ISEs. We did so by first aggregating ISE scores and ratings at a company level. The aggregation reduces the endogenous association between company ratings and ISEs in individual reviews. We then conducted an OLS regression using our two main sustainability facets as predictors (`staff welfare' and `financial benefits') while also controlling for a company's total number of reviews. As reported in Table~\ref{tab:company_rating_regressions_pca}, we found that these two sustainability facets could explain up to 64\% of the variance in a company's ratings; particularly noteworthy was that the staff welfare facet of corporate internal sustainability was strongly positively correlated with all aspects of a company's success, including balance and culture, in line with previous research findings~\cite{rao2017work,isensee2020relationship}.\\

\begin{table}[!tbp] \centering
\scriptsize
\begin{tabular}{@{\extracolsep{5pt}}lccccc}
\\[-1.8ex]\hline
\hline \\[-1.8ex]
\\[-1.8ex] & Balance & Career & Culture & Management & Overall \\
\hline \\[-1.8ex]
Const & 16.465$^{**}$ & 29.995$^{***}$ & 30.165$^{***}$ & 24.863$^{***}$ & 5.651$^{}$ \\
  & (6.286) & (4.723) & (5.990) & (5.614) & (4.740) \\
 Staff Welfare (PC1) & 0.732$^{***}$ & 0.681$^{***}$ & 0.721$^{***}$ & 0.615$^{***}$ & 0.766$^{***}$ \\
  & (0.095) & (0.071) & (0.090) & (0.085) & (0.067) \\
 Financial Benefits (PC2) & 0.183$^{*}$ & -0.164$^{**}$ & -0.160$^{}$ & -0.201$^{**}$ & 0.313$^{***}$ \\
  & (0.109) & (0.082) & (0.104) & (0.097) & (0.076) \\
 Total Reviews & & & & & 0.102$^{}$ \\
  & & & & & (0.063) \\
\hline \\[-1.8ex]
 Observations & 84 & 84 & 84 & 84 & 84 \\
 $R^2$ & 0.443 & 0.534 & 0.444 & 0.405 & 0.658 \\
 Adjusted $R^2$ & 0.429 & 0.523 & 0.430 & 0.390 & 0.645 \\
 Residual Std. Error & 16.283(df = 81) & 12.234(df = 81) & 15.515(df = 81) & 14.541(df = 81) & 11.355(df = 80)  \\
 F Statistic & 32.174$^{***}$ & 46.461$^{***}$ & 32.289$^{***}$ & 27.518$^{***}$ & 51.319$^{***}$ \\
 & (df = 2.0; 81.0) & (df = 2.0; 81.0) & (df = 2.0; 81.0) & (df = 2.0; 81.0) & (df = 3.0; 80.0) \\
\hline
\hline \\[-1.8ex]
\textit{Note:} & \multicolumn{5}{r}{$^{*}$p$<$0.1; $^{**}$p$<$0.05; $^{***}$p$<$0.01} \\
\end{tabular}
\caption{Predicting company online ratings from the two main facets of sustainability (staff welfare and financial benefits) using a stepAIC analysis on an OLS regression.}
\label{tab:company_rating_regressions_pca}
\end{table}

\begin{figure}[t]
\centering
\includegraphics[width=\linewidth]{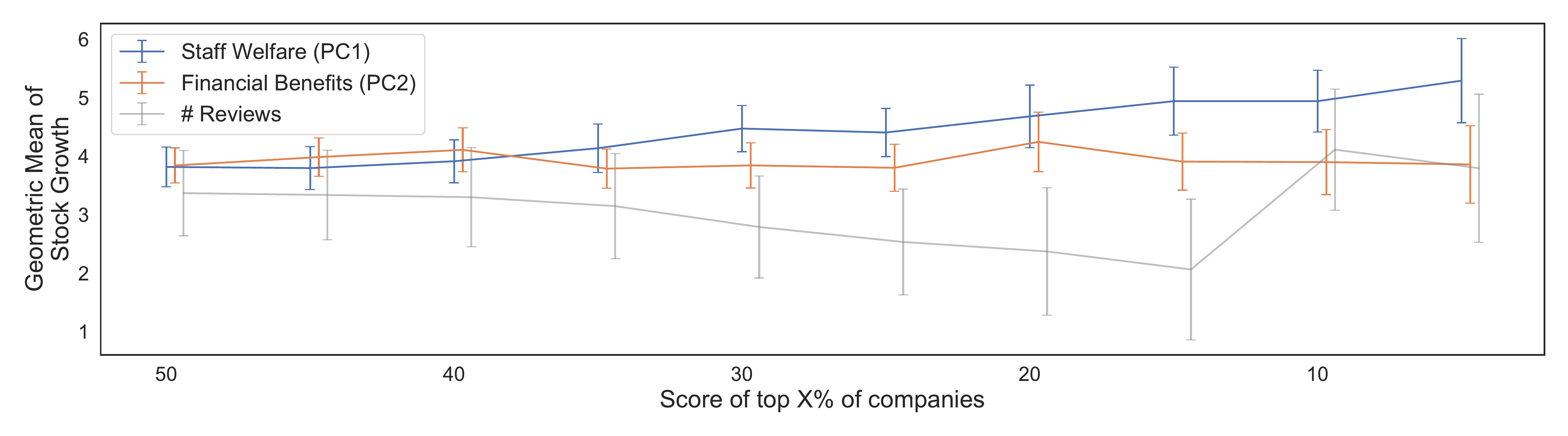}
\caption{Geometric mean of stock growth values for increasing ISE score ranking.}
\label{fig:gm_bin_plot}
\end{figure}

\noindent
\textit{Sustainability and company stock growth.} We obtained stock data for 84 of the 104 companies in our dataset, from 2009 to 2019, using the Yahoo Finance portal~\cite{WinNT}. For each company, we calculated the geometric mean of its stock growth during such period; we used the geometric mean since the distribution of stock growth values across companies was heavy-tailed (as reported in  Supplementary Information). To inspect whether a company's financial success (measured as stock growth) was associated with its sustainability efforts, we then plotted in Figure~\ref{fig:gm_bin_plot} the geometric mean of its stock growth ($y$ axis) against its ranking in terms of the staff welfare facet of sustainability and the financial benefits facet of sustainability ($x$ axis). We also included in the figure the total number of reviews, to check whether stock growth was merely associated with the company's popularity rather than its internal sustainability practices.

As showed in Figure~\ref{fig:gm_bin_plot}, companies that focused on both staff welfare and financial benefits sustainability tended to have high stock growth; between the two facets, it was staff welfare that most strongly associated with high stock growth, in line with previous research~\cite{bcg}. 
Notably, companies with high stock growth did not invest as heavily in financial sustainability only, bolstering previous work which noted that focusing on staff welfare sustainability could lead to greater stakeholder engagement even without high pay~\cite{ziegler2007effect}. Overall, our results suggest that a company's financial success is associated with its investment in internal sustainability practices, but only if they focus on a holistic approach to sustainability that tackles both staff welfare and financial benfits.

\subsection*{RQ3: Is sustainability associated with specific industry sectors?} 

\begin{figure}[!htb]
\begin{center}
\includegraphics[width=\linewidth]{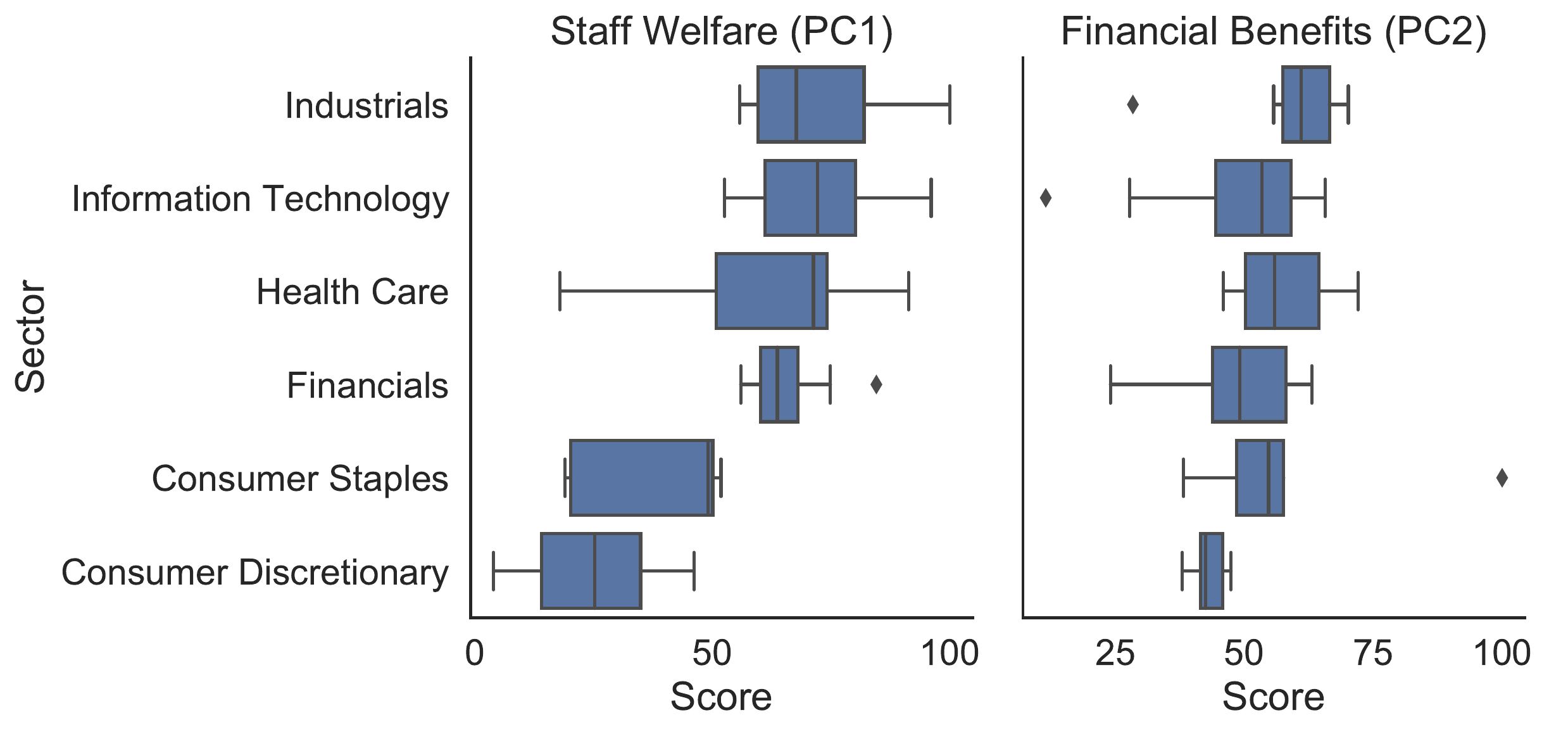}
\end{center}
\caption{\textbf{Sustainability and industry sector}. Boxplots showing the distribution of the staff welfare and financial benefits ISE scores across different industry sectors. 
}
\label{fig:sector_analysis}
\end{figure}

\noindent To examine whether certain industry sectors were leading the corporate sustainability agenda, we plotted in  Figure~\ref{fig:sector_analysis} the distribution of the two facets of sustainability for each industry sector. We further conducted a MANOVA analysis and found the differences in sustainability scores to be significant across sectors. In terms of staff welfare sustainability efforts, we found Industrials and IT to lead, possibly due to recent investment in this type of sustainability initiatives~\cite{higon2017ict}. The Financial sector followed, while the health care one exhibited very high variability. This could be explained by health care professionals often sacrificing personal well-being and work-life balance due to the highly demanding nature of their work~\cite{schwartz2019work,shanafelt2015changes}. We found consumer staples and consumer discretionary to lag significantly behind. This was also the case when looking at financial benefits, although differences between sectors were smaller along this facet of internal sustainability efforts.

\begin{figure}[t]
\centering
\includegraphics[width=\linewidth]{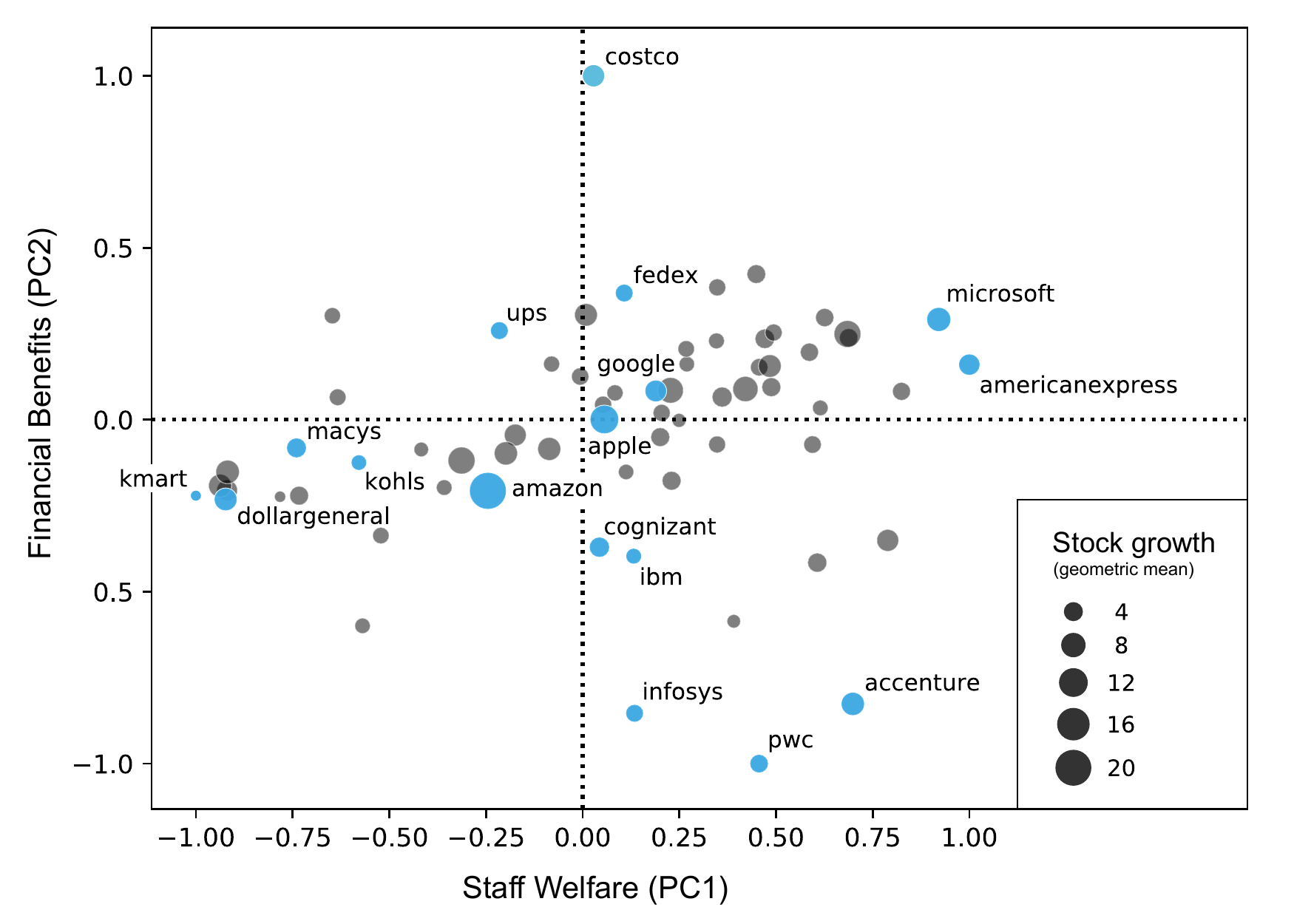}
\caption{\textbf{Scatterplot of the scores of each company's staff welfare \emph{vs.} financial benefits.} The size of a company's dot represents its stock growth. We highlighted in blue some of the companies to assess them qualitatively. Consumer staples and discretionary companies like Kmart, Macy's, and Kohl's scored low for both types of sustainability. Traditional IT companies like Infosys, IBM, and Accenture scored high for staff welfare sustainability but not for financial benefits sustainability.}
\label{fig:pca_quadrant}
\end{figure}

Figure~\ref{fig:sector_analysis} offered an overview of engagement with sustainability efforts at industry sector level. To reveal more nuanced variations {\em within} the same sector, we plotted individual companies' engagement with each of the two main sustainability facets in Figure~\ref{fig:pca_quadrant}. Notable variations emerged: our sector-based analysis revealed low sustainability scores for consumer discretionary and staples companies overall; upon closer inspection, we found some companies (e.g., Dollar General, K-mart) to indeed score low on both facets of ISEs, while others (e.g., Costco) to score low on staff welfare but high in the financial benefits facet of sustainability, a phenomenon noted in previous work too~\cite{cascio2006decency}. Variations emerged also within the IT sector, previously showed to be leading sustainability efforts on both dimensions: a more nuanced investigation revealed high sustainability scores on both financial benefits and staff welfare ISEs for companies like Microsoft, Google, and Apple; however, more traditional IT companies like Infosys, IBM, and Cognizant scored high on staff welfare sustainability only. 

One must be mindful that comparisons among (same-sector) companies were further affected by the type of employees reviewing their employer. In our study, this was apparent for companies like Amazon, that enjoyed high stock growth but surprisingly scored low for both types of sustainability. Despite the company employing a large number of software engineers as well as warehouse workers, upon close inspection, we found the most common roles of Amazon employees in our reviews to be `warehouse associate' and `warehouse worker'. Previous research did find logistic workers at Amazon to face poor working conditions~\cite{amzn,chan2015examining}, corroborating the low sustainability scores that our method computed for this company. Furthermore, previous literature noted that Amazon's lack of focus on sustainability practices has yet to hurt its profitability  \cite{amzn,chan2015examining}.

\section*{Discussion}

Many companies are under constant pressure to invest in a wide range of internal sustainability practices designed to enhance working conditions~\cite{barko2022shareholder,jakob2022like}. However, the benefits of such investments for both the company and its stakeholders are often difficult to assess. By examining how employees form perceptions of their company's engagement with ISEs, this research spells out the microfoundations of internal sustainability and provides evidence of the strategic importance of investing in business practices and policies geared towards ISEs~\cite{de2021reimagining,zhao2022influence}. 

By examining how the wider UN SDGs agenda can be translated into diverse internal corporate efforts directed towards employees, our work offers substantive methodological, conceptual, and empirical contributions to internal sustainability research and managerial practice. More specifically, it offers two main theoretical contributions. The first has to do with the conceptualization of ISEs.  We have shown  how the sustainability agenda brought forward by the introduction of the UN SDGs informs and shapes six  sustainability efforts within a company. Efforts to do with health, education, diversity, monetary benefits, supporting infrastructure, and a supportive atmosphere. While the existing literature often presents sustainability as a monolithic construct~\cite{chen2020corporate,liu2020too}, our two-factor conceptualization of ISEs delineated the two core strategic aspects that companies should carefully balance when implementing ISEs: one aspect had to do with traditional financial benefits (e.g., salary, bonuses), and the other had to do with broader aspects of staff welfare (e.g., diversity, atmosphere). The second theoretical implication enhances the understanding of what makes companies economically successful and how internal sustainability practices differ by sector, especially in emerging sectors like IT.

This work also offers practical implications, and it does so for  three main stakeholders. The first stakeholder consists of scholars. Our method is grounded in the UN SDGs and performed consistently well across several rounds of external validation.  By providing a robust framework for examining mentions of ISEs through  automated text analysis, new textual datasets could be academically studied in the future. 

The second stakeholder consists of policy makers. We showed that high levels of ISEs engagement (not only for financial aspects but also for general staff welfare) were associated with high economic growth. This result supports policies in recent years that have fostered a corporate culture that goes beyond financial rewards and are oriented towards equality and well-being~\cite{triana2019perceived}. Beyond company efforts, policy makers themselves would be able to strategically decide which ISEs to incentivize with taxation schemes or set out a legislation agenda that would attract workers who care about specific ISEs. To inform more targeted interventions, we also showed that the impact of engaging with ISEs varies across  sectors: companies in the IT and business-to-business industrial goods sectors outperformed companies that produce and commercialize consumer goods. This finding is noteworthy as previous research shows that sustainability signals tend to be stronger in business-to-consumer than in business-to-business market contexts (c.f., \cite{hoejmose2012green})

The third stakeholder consists of company managers. By reflecting employees’ perceptions, our analytical framework represents an invaluable tool to operationalize the microfoundations of internal sustainability, assess how corporate efforts in this area directly impact employees, and  quantify and qualify the extent to which corporate engagement with ISEs becomes visible to employees across different organizational levels.

Our work comes with five main limitations though. The first is that our list of ISEs may not be accurate or exhaustive. While the corporate sustainability literature has focused on initiatives that are external to a company and have an impact on the wider world's sustainability, the practices that are internal to a company and have an impact on employees received less attention. As a result we only found non-comprehensive frameworks for internal sustainability practices suggested, such as those focusing on social aspects only \cite{baumgartner2010corporate}. To tackle that, we started from the well-grounded definitions of the UN  sustainability goals, used a principled mixed-method approach to paraphrase those most relevant to the corporate context, and validated the resulting list with both qualitative and quantitative approaches. These approaches are generalizable, in that they could be used to study other constructs appearing in reviews in the future (e.g., how employees in a company deal with stress).

The second limitation is that, since the reviewing site was founded in 2008, key financial events prior 2008 (e.g., the dot-com bubble in the late 1990s) may have impacted our results but could not be accounted for because of lack of data.

The third limitation is that the number of companies under study is invariably limited. We were able to study 104 major companies, largely because the other companies had a limited number of reviews that did not allow for automatic processing. Future work should explore alternative mixed-method approaches (likely qualitative ones) to study ISEs for these companies.

The fourth limitation is the lack of causal claims. Given our data, we could not assess the causal direction between ISEs and socio-economic returns. More specifically, we could not assess whether focusing on ISEs led to better socio-economic returns (e.g., stock growth), whether better socio-economic conditions created a breeding ground for fostering ISEs, or whether these two causal relations were in a self-reinforcing cycle.

The fifth and final limitation has to do with the representativeness  of our data. Companies in certain sectors (e.g., IT) may have been reviewed more often than  those in other sectors (e.g., consumer discretionary). Despite that possibility, in Supplementary Information, we show that our data was still representative along three major dimensions: 
a) the distribution of industry sectors of the S\&P 500 companies, which our data matched without over-representing any specific sector; b) official population in a state, which scaled linearly with the number of employees in the state in our data; and c) number of company headquarters in a state from official sources, which has a nearly perfect correlation with the number of headquarters per state in our data. Finally, despite the platform's mechanisms to guarantee review quality, as discussed in the section \emph{Data}, we acknowledge that potential self-selection bias could cause our reviewers' sample to be non-representative. To reduce the impact of such a bias and ensure robustness, we restricted our analyses to companies having at least 1000 reviews.

\section*{Data availability}

We made our code and data available in a readily usable format (\url{http://social-dynamics.net/InsiderStories/}) for reproducibility. For each company, we shared the following attributes: \texttt{company name}, \texttt{\#total reviews}, \texttt{\#ISE reviews}, \texttt{average rating}, \texttt{rating of work-life balance}, \texttt{rating of career prospects}, \texttt{rating of the company}, \texttt{rating of the culture}, \texttt{rating of the management}, \texttt{ISE type}, \texttt{stock values/growth for: 2009, 2012, 2014, 2019}, and \texttt{industry sector}.

\bibliography{main}

\begin{thebibliography}{10}
\urlstyle{rm}
\expandafter\ifx\csname url\endcsname\relax
  \def\url#1{\texttt{#1}}\fi
\expandafter\ifx\csname urlprefix\endcsname\relax\def\urlprefix{URL }\fi
\expandafter\ifx\csname doiprefix\endcsname\relax\def\doiprefix{DOI: }\fi
\providecommand{\bibinfo}[2]{#2}
\providecommand{\eprint}[2][]{\url{#2}}

\bibitem{bai2020supply}
\bibinfo{author}{Bai, C.} \& \bibinfo{author}{Sarkis, J.}
\newblock \bibinfo{journal}{\bibinfo{title}{A supply chain transparency and
  sustainability technology appraisal model for blockchain technology}}.
\newblock {\emph{\JournalTitle{International Journal of Production Research}}}
  \textbf{\bibinfo{volume}{58}}, \bibinfo{pages}{2142--2162}
  (\bibinfo{year}{2020}).

\bibitem{de2021reimagining}
\bibinfo{author}{de~Ruyter, K.} \emph{et~al.}
\newblock \bibinfo{title}{Reimagining marketing strategy: driving the debate on
  grand challenges} (\bibinfo{year}{2021}).

\bibitem{serafeim2020social}
\bibinfo{author}{Serafeim, G.}
\newblock \bibinfo{journal}{\bibinfo{title}{Social-impact efforts that create
  real value}}.
\newblock {\emph{\JournalTitle{Harvard Business Review}}}
  \textbf{\bibinfo{volume}{98}}, \bibinfo{pages}{38--48}
  (\bibinfo{year}{2020}).

\bibitem{wang2022social}
\bibinfo{author}{Wang, H.}, \bibinfo{author}{Jia, M.}, \bibinfo{author}{Xiang,
  Y.} \& \bibinfo{author}{Lan, Y.}
\newblock \bibinfo{journal}{\bibinfo{title}{Social performance feedback and
  firm communication strategy}}.
\newblock {\emph{\JournalTitle{Journal of Management}}}
  \textbf{\bibinfo{volume}{48}}, \bibinfo{pages}{2382--2420}
  (\bibinfo{year}{2022}).

\bibitem{doi:10.1177/0022242921992052}
\bibinfo{author}{Gonzalez-Arcos, C.}, \bibinfo{author}{Joubert, A.~M.},
  \bibinfo{author}{Scaraboto, D.}, \bibinfo{author}{Guesalaga, R.} \&
  \bibinfo{author}{Sandberg, J.}
\newblock \bibinfo{journal}{\bibinfo{title}{“how do i carry all this now?”
  understanding consumer resistance to sustainability interventions}}.
\newblock {\emph{\JournalTitle{Journal of Marketing}}}
  \textbf{\bibinfo{volume}{85}}, \bibinfo{pages}{44--61},
  \doiprefix\url{10.1177/0022242921992052} (\bibinfo{year}{2021}).
\newblock \eprint{https://doi.org/10.1177/0022242921992052}.

\bibitem{martin2021exploring}
\bibinfo{author}{Mart{\'\i}n-de Castro, G.}
\newblock \bibinfo{title}{Exploring the market side of corporate
  environmentalism: Reputation, legitimacy and stakeholders' engagement}
  (\bibinfo{year}{2021}).

\bibitem{paine2014sustainability}
\bibinfo{author}{Paine, L.~S.}
\newblock \bibinfo{journal}{\bibinfo{title}{Sustainability in the boardroom}}.
\newblock {\emph{\JournalTitle{Harvard Business Review}}}
  \textbf{\bibinfo{volume}{92}}, \bibinfo{pages}{86--94}
  (\bibinfo{year}{2014}).

\bibitem{chatzopoulou2022corporate}
\bibinfo{author}{Chatzopoulou, E.-C.}, \bibinfo{author}{Manolopoulos, D.} \&
  \bibinfo{author}{Agapitou, V.}
\newblock \bibinfo{journal}{\bibinfo{title}{Corporate social responsibility and
  employee outcomes: Interrelations of external and internal orientations with
  job satisfaction and organizational commitment}}.
\newblock {\emph{\JournalTitle{Journal of Business Ethics}}}
  \textbf{\bibinfo{volume}{179}}, \bibinfo{pages}{795--817}
  (\bibinfo{year}{2022}).

\bibitem{kelliher2019all}
\bibinfo{author}{Kelliher, C.}, \bibinfo{author}{Richardson, J.} \&
  \bibinfo{author}{Boiarintseva, G.}
\newblock \bibinfo{journal}{\bibinfo{title}{All of work? all of life?
  reconceptualising work-life balance for the 21st century}}.
\newblock {\emph{\JournalTitle{Human Resource Management Journal}}}
  \textbf{\bibinfo{volume}{29}}, \bibinfo{pages}{97--112}
  (\bibinfo{year}{2019}).

\bibitem{NADEEM2017874}
\bibinfo{author}{Nadeem, M.}, \bibinfo{author}{Zaman, R.} \&
  \bibinfo{author}{Saleem, I.}
\newblock \bibinfo{journal}{\bibinfo{title}{Boardroom gender diversity and
  corporate sustainability practices: Evidence from australian securities
  exchange listed firms}}.
\newblock {\emph{\JournalTitle{Journal of Cleaner Production}}}
  \textbf{\bibinfo{volume}{149}}, \bibinfo{pages}{874--885},
  \doiprefix\url{https://doi.org/10.1016/j.jclepro.2017.02.141}
  (\bibinfo{year}{2017}).

\bibitem{cassino2019race}
\bibinfo{author}{Cassino, D.} \& \bibinfo{author}{Besen-Cassino, Y.}
\newblock \bibinfo{journal}{\bibinfo{title}{Race, threat and workplace sexual
  harassment: The dynamics of harassment in the united states, 1997--2016}}.
\newblock {\emph{\JournalTitle{Gender, Work and Organization}}}
  \textbf{\bibinfo{volume}{26}}, \bibinfo{pages}{1221--1240}
  (\bibinfo{year}{2019}).

\bibitem{giauque2019stress}
\bibinfo{author}{Giauque, D.}, \bibinfo{author}{Anderfuhren-Biget, S.} \&
  \bibinfo{author}{Varone, F.}
\newblock \bibinfo{journal}{\bibinfo{title}{Stress and turnover intents in
  international organizations: social support and work--life balance as
  resources}}.
\newblock {\emph{\JournalTitle{The International Journal of Human Resource
  Management}}} \textbf{\bibinfo{volume}{30}}, \bibinfo{pages}{879--901}
  (\bibinfo{year}{2019}).

\bibitem{wang2012explaining}
\bibinfo{author}{Wang, J.} \& \bibinfo{author}{Verma, A.}
\newblock \bibinfo{journal}{\bibinfo{title}{Explaining organizational
  responsiveness to work-life balance issues: The role of business strategy and
  high-performance work systems}}.
\newblock {\emph{\JournalTitle{Human Resource Management}}}
  \textbf{\bibinfo{volume}{51}}, \bibinfo{pages}{407--432}
  (\bibinfo{year}{2012}).

\bibitem{peloza2011how}
\bibinfo{author}{Peloza, J.} \& \bibinfo{author}{Shang, J.}
\newblock \bibinfo{journal}{\bibinfo{title}{How can corporate social
  responsibility activities create value for stakeholders? a systematic
  review.}}
\newblock {\emph{\JournalTitle{Journal of the Academy of Marketing Science}}}
  (\bibinfo{year}{2011}).

\bibitem{wced1987world}
\bibinfo{author}{WCED, S. W.~S.}
\newblock \bibinfo{journal}{\bibinfo{title}{World commission on environment and
  development}}.
\newblock {\emph{\JournalTitle{Our common future}}}
  \textbf{\bibinfo{volume}{17}}, \bibinfo{pages}{1--91} (\bibinfo{year}{1987}).

\bibitem{nations2015transforming}
\bibinfo{author}{Nations, U.}
\newblock \bibinfo{journal}{\bibinfo{title}{Transforming our world: The 2030
  agenda for sustainable development}}.
\newblock {\emph{\JournalTitle{New York: United Nations, Department of Economic
  and Social Affairs}}}  (\bibinfo{year}{2015}).

\bibitem{das2020modeling}
\bibinfo{author}{Das~Swain, V.} \emph{et~al.}
\newblock \bibinfo{title}{Modeling organizational culture with workplace
  experiences shared on glassdoor}.
\newblock In \emph{\bibinfo{booktitle}{Proceedings of the 2020 CHI conference
  on human factors in computing systems}}, \bibinfo{pages}{1--15}
  (\bibinfo{year}{2020}).

\bibitem{WinNT}
\bibinfo{title}{{Yahoo Finance portal}}.
\newblock \bibinfo{howpublished}{\url{https://ﬁnance.yahoo.com.}}
\newblock \bibinfo{note}{Accessed: 2021-08-02}.

\bibitem{elo2008qualitative}
\bibinfo{author}{Elo, S.} \& \bibinfo{author}{Kyng{\"a}s, H.}
\newblock \bibinfo{journal}{\bibinfo{title}{The qualitative content analysis
  process}}.
\newblock {\emph{\JournalTitle{Journal of advanced nursing}}}
  \textbf{\bibinfo{volume}{62}}, \bibinfo{pages}{107--115}
  (\bibinfo{year}{2008}).

\bibitem{reimers2019sentence}
\bibinfo{author}{Reimers, N.} \& \bibinfo{author}{Gurevych, I.}
\newblock \bibinfo{title}{Sentence-bert: Sentence embeddings using siamese
  bert-networks}.
\newblock In \emph{\bibinfo{booktitle}{Proceedings of the 2019 Conference on
  Empirical Methods in Natural Language Processing and the 9th International
  Joint Conference on Natural Language Processing (EMNLP-IJCNLP)}},
  \bibinfo{pages}{3982--3992} (\bibinfo{year}{2019}).

\bibitem{choi20ten}
\bibinfo{author}{Choi, M.}, \bibinfo{author}{Aiello, L.~M.},
  \bibinfo{author}{Varga, K.~Z.} \& \bibinfo{author}{Quercia, D.}
\newblock \bibinfo{title}{Ten social dimensions of conversations and
  relationships}.
\newblock In \emph{\bibinfo{booktitle}{{Proceedings of The ACM Web Conference
  (WWW)}}}, \bibinfo{pages}{1514–1525},
  \doiprefix\url{10.1145/3366423.3380224} (\bibinfo{year}{2020}).

\bibitem{denzin2012triangulation}
\bibinfo{author}{Denzin, N.~K.}
\newblock \bibinfo{journal}{\bibinfo{title}{Triangulation 2.0}}.
\newblock {\emph{\JournalTitle{Journal of mixed methods research}}}
  \textbf{\bibinfo{volume}{6}}, \bibinfo{pages}{80--88} (\bibinfo{year}{2012}).

\bibitem{chung2018flexible}
\bibinfo{author}{Chung, H.} \& \bibinfo{author}{Van~der Lippe, T.}
\newblock \bibinfo{journal}{\bibinfo{title}{Flexible working, work--life
  balance, and gender equality: Introduction}}.
\newblock {\emph{\JournalTitle{Social Indicators Research}}}
  \bibinfo{pages}{1--17} (\bibinfo{year}{2018}).

\bibitem{lyonette2015part}
\bibinfo{author}{Lyonette, C.}
\newblock \bibinfo{journal}{\bibinfo{title}{Part-time work, work--life balance
  and gender equality}}.
\newblock {\emph{\JournalTitle{Journal of Social Welfare and Family Law}}}
  \textbf{\bibinfo{volume}{37}}, \bibinfo{pages}{321--333}
  (\bibinfo{year}{2015}).

\bibitem{rao2017work}
\bibinfo{author}{Rao, I.}
\newblock \bibinfo{journal}{\bibinfo{title}{Work-life balance for sustainable
  human development: Cultural intelligence as enabler}}.
\newblock {\emph{\JournalTitle{Journal of Human Behavior in the Social
  Environment}}} \textbf{\bibinfo{volume}{27}}, \bibinfo{pages}{706--713}
  (\bibinfo{year}{2017}).

\bibitem{isensee2020relationship}
\bibinfo{author}{Isensee, C.}, \bibinfo{author}{Teuteberg, F.},
  \bibinfo{author}{Griese, K.-M.} \& \bibinfo{author}{Topi, C.}
\newblock \bibinfo{journal}{\bibinfo{title}{The relationship between
  organizational culture, sustainability, and digitalization in smes: A
  systematic review}}.
\newblock {\emph{\JournalTitle{Journal of Cleaner Production}}}
  \bibinfo{pages}{122944} (\bibinfo{year}{2020}).

\bibitem{bcg}
\bibinfo{title}{{Diversity, Equity, and Inclusion Still Matter in a Pandemic}}.
\newblock
  \bibinfo{howpublished}{\url{https://www.bcg.com/publications/2020/value-of-investing-in-diversity-equity-and-inclusion-during-a-pandemic}}.
\newblock \bibinfo{note}{Accessed: 2021-08-06}.

\bibitem{ziegler2007effect}
\bibinfo{author}{Ziegler, A.}, \bibinfo{author}{Schr{\"o}der, M.} \&
  \bibinfo{author}{Rennings, K.}
\newblock \bibinfo{journal}{\bibinfo{title}{The effect of environmental and
  social performance on the stock performance of european corporations}}.
\newblock {\emph{\JournalTitle{Environmental and Resource Economics}}}
  \textbf{\bibinfo{volume}{37}}, \bibinfo{pages}{661--680}
  (\bibinfo{year}{2007}).

\bibitem{higon2017ict}
\bibinfo{author}{Hig{\'o}n, D.~A.}, \bibinfo{author}{Gholami, R.} \&
  \bibinfo{author}{Shirazi, F.}
\newblock \bibinfo{journal}{\bibinfo{title}{Ict and environmental
  sustainability: A global perspective}}.
\newblock {\emph{\JournalTitle{Telematics and Informatics}}}
  \textbf{\bibinfo{volume}{34}}, \bibinfo{pages}{85--95}
  (\bibinfo{year}{2017}).

\bibitem{schwartz2019work}
\bibinfo{author}{Schwartz, S.~P.} \emph{et~al.}
\newblock \bibinfo{journal}{\bibinfo{title}{Work-life balance behaviours
  cluster in work settings and relate to burnout and safety culture: a
  cross-sectional survey analysis}}.
\newblock {\emph{\JournalTitle{BMJ Quality and Safety}}}
  \textbf{\bibinfo{volume}{28}}, \bibinfo{pages}{142--150}
  (\bibinfo{year}{2019}).

\bibitem{shanafelt2015changes}
\bibinfo{author}{Shanafelt, T.~D.} \emph{et~al.}
\newblock \bibinfo{title}{Changes in burnout and satisfaction with work-life
  balance in physicians and the general us working population between 2011 and
  2014}.
\newblock In \emph{\bibinfo{booktitle}{Mayo clinic proceedings}},
  vol.~\bibinfo{volume}{90}, \bibinfo{pages}{1600--1613}
  (\bibinfo{organization}{Elsevier}, \bibinfo{year}{2015}).

\bibitem{cascio2006decency}
\bibinfo{author}{Cascio, W.~F.}
\newblock \bibinfo{journal}{\bibinfo{title}{Decency means more than “always
  low prices”: A comparison of costco to wal-mart's sam's club}}.
\newblock {\emph{\JournalTitle{Academy of Management perspectives}}}
  \textbf{\bibinfo{volume}{20}}, \bibinfo{pages}{26--37}
  (\bibinfo{year}{2006}).

\bibitem{amzn}
\bibinfo{title}{{Amazon's no show on sustainability}}.
\newblock
  \bibinfo{howpublished}{\url{https://www.theguardian.com/sustainable-business/amazon}}.
\newblock \bibinfo{note}{Accessed: 2021-08-06}.

\bibitem{chan2015examining}
\bibinfo{author}{Chan, I.}
\newblock \emph{\bibinfo{title}{Examining the cost of Amazon. com’s success
  using the triple bottom line}}.
\newblock Master's thesis, \bibinfo{school}{Humboldt State University}
  (\bibinfo{year}{2015}).

\bibitem{barko2022shareholder}
\bibinfo{author}{Barko, T.}, \bibinfo{author}{Cremers, M.} \&
  \bibinfo{author}{Renneboog, L.}
\newblock \bibinfo{journal}{\bibinfo{title}{Shareholder engagement on
  environmental, social, and governance performance}}.
\newblock {\emph{\JournalTitle{Journal of Business Ethics}}}
  \textbf{\bibinfo{volume}{180}}, \bibinfo{pages}{777--812}
  (\bibinfo{year}{2022}).

\bibitem{jakob2022like}
\bibinfo{author}{Jakob, E.~A.}, \bibinfo{author}{Steinmetz, H.},
  \bibinfo{author}{Wehner, M.~C.}, \bibinfo{author}{Engelhardt, C.} \&
  \bibinfo{author}{Kabst, R.}
\newblock \bibinfo{journal}{\bibinfo{title}{Like it or not: when corporate
  social responsibility does not attract potential applicants}}.
\newblock {\emph{\JournalTitle{Journal of Business Ethics}}}
  \textbf{\bibinfo{volume}{178}}, \bibinfo{pages}{105--127}
  (\bibinfo{year}{2022}).

\bibitem{zhao2022influence}
\bibinfo{author}{Zhao, X.}, \bibinfo{author}{Wu, C.}, \bibinfo{author}{Chen,
  C.~C.} \& \bibinfo{author}{Zhou, Z.}
\newblock \bibinfo{journal}{\bibinfo{title}{The influence of corporate social
  responsibility on incumbent employees: A meta-analytic investigation of the
  mediating and moderating mechanisms}}.
\newblock {\emph{\JournalTitle{Journal of Management}}}
  \textbf{\bibinfo{volume}{48}}, \bibinfo{pages}{114--146}
  (\bibinfo{year}{2022}).

\bibitem{chen2020corporate}
\bibinfo{author}{Chen, Z.}, \bibinfo{author}{Hang, H.},
  \bibinfo{author}{Pavelin, S.} \& \bibinfo{author}{Porter, L.}
\newblock \bibinfo{journal}{\bibinfo{title}{Corporate social (ir)
  responsibility and corporate hypocrisy: Warmth, motive and the protective
  value of corporate social responsibility}}.
\newblock {\emph{\JournalTitle{Business Ethics Quarterly}}}
  \textbf{\bibinfo{volume}{30}}, \bibinfo{pages}{486--524}
  (\bibinfo{year}{2020}).

\bibitem{liu2020too}
\bibinfo{author}{Liu, A.~Z.}, \bibinfo{author}{Liu, A.~X.},
  \bibinfo{author}{Wang, R.} \& \bibinfo{author}{Xu, S.~X.}
\newblock \bibinfo{journal}{\bibinfo{title}{Too much of a good thing? the
  boomerang effect of firms’ investments on corporate social responsibility
  during product recalls}}.
\newblock {\emph{\JournalTitle{Journal of Management Studies}}}
  \textbf{\bibinfo{volume}{57}}, \bibinfo{pages}{1437--1472}
  (\bibinfo{year}{2020}).

\bibitem{triana2019perceived}
\bibinfo{author}{Triana, M. d.~C.}, \bibinfo{author}{Jayasinghe, M.},
  \bibinfo{author}{Pieper, J.~R.}, \bibinfo{author}{Delgado, D.~M.} \&
  \bibinfo{author}{Li, M.}
\newblock \bibinfo{journal}{\bibinfo{title}{Perceived workplace gender
  discrimination and employee consequences: A meta-analysis and complementary
  studies considering country context}}.
\newblock {\emph{\JournalTitle{Journal of management}}}
  \textbf{\bibinfo{volume}{45}}, \bibinfo{pages}{2419--2447}
  (\bibinfo{year}{2019}).

\bibitem{hoejmose2012green}
\bibinfo{author}{Hoejmose, S.}, \bibinfo{author}{Brammer, S.} \&
  \bibinfo{author}{Millington, A.}
\newblock \bibinfo{journal}{\bibinfo{title}{“green” supply chain
  management: The role of trust and top management in b2b and b2c markets}}.
\newblock {\emph{\JournalTitle{Industrial Marketing Management}}}
  \textbf{\bibinfo{volume}{41}}, \bibinfo{pages}{609--620}
  (\bibinfo{year}{2012}).

\bibitem{baumgartner2010corporate}
\bibinfo{author}{Baumgartner, R.~J.} \& \bibinfo{author}{Ebner, D.}
\newblock \bibinfo{journal}{\bibinfo{title}{Corporate sustainability
  strategies: sustainability profiles and maturity levels}}.
\newblock {\emph{\JournalTitle{Sustainable development}}}
  \textbf{\bibinfo{volume}{18}}, \bibinfo{pages}{76--89}
  (\bibinfo{year}{2010}).

\bibitem{devlin2018bert}
\bibinfo{author}{Devlin, J.}, \bibinfo{author}{Chang, M.-W.},
  \bibinfo{author}{Lee, K.} \& \bibinfo{author}{Toutanova, K.}
\newblock \bibinfo{journal}{\bibinfo{title}{Bert: Pre-training of deep
  bidirectional transformers for language understanding}}.
\newblock {\emph{\JournalTitle{arXiv preprint arXiv:1810.04805}}}
  (\bibinfo{year}{2018}).

\bibitem{morgan2017employee}
\bibinfo{author}{Morgan, J.}
\newblock \emph{\bibinfo{title}{The employee experience advantage: How to win
  the war for talent by giving employees the workspaces they want, the tools
  they need, and a culture they can celebrate}} (\bibinfo{publisher}{John Wiley
  and Sons}, \bibinfo{year}{2017}).

\bibitem{burritt2010sustainability}
\bibinfo{author}{Burritt, R.~L.} \& \bibinfo{author}{Schaltegger, S.}
\newblock \bibinfo{journal}{\bibinfo{title}{Sustainability accounting and
  reporting: fad or trend?}}
\newblock {\emph{\JournalTitle{Accounting, Auditing and Accountability
  Journal}}}  (\bibinfo{year}{2010}).

\bibitem{equileap}
\bibinfo{title}{{GENDER EQUALITY IN THE U.S.} assessing 500 leading companies
  on workplace equality including healthcare benefits}.
\newblock
  \bibinfo{howpublished}{\url{https://equileap.com/wp-content/uploads/2020/12/Equileap_US_Report_2020.pdf
  }}.
\newblock \bibinfo{note}{Accessed: 2021-08-06}.

\bibitem{fti}
\bibinfo{title}{{The Fashion Transparency Index 2021}}.
\newblock
  \bibinfo{howpublished}{\url{https://www.fashionrevolution.org/about/transparency/}}.
\newblock \bibinfo{note}{Accessed: 2021-08-06}.

\bibitem{webber2010similarity}
\bibinfo{author}{Webber, W.}, \bibinfo{author}{Moffat, A.} \&
  \bibinfo{author}{Zobel, J.}
\newblock \bibinfo{journal}{\bibinfo{title}{A similarity measure for indefinite
  rankings}}.
\newblock {\emph{\JournalTitle{ACM Transactions on Information Systems
  (TOIS)}}} \textbf{\bibinfo{volume}{28}}, \bibinfo{pages}{1--38}
  (\bibinfo{year}{2010}).

\end{thebibliography}

\section*{Ethical approval and informed consent statement}
This article does not contain any studies with human participants performed by any of the authors.

\section*{Contributions of the authors}
All authors contributed to the research design and writing of the paper. IS was mainly responsible for the linguistics and statistical analysis with guidance from SS, LC, and DQ. SS cleaned, collected and pre-processed the data. MM provided theoretical guidance. All authors wrote, read, and approved the final manuscript.

\section*{Competing interests}
The authors declare no competing interests.

\section*{Acknowledgements}
We thank Dr. Edyta Bogucka for her help with some of the figures in this manuscript.

\section*{Supplementary Information}
\appendix

\section{Details of the dataset}
We collected a total of 713,018 reviews published by current and former employees on a popular company reviewing site from the start of 2008 up until the first quarter of 2020. We filtered out reviews belonging to non-US based companies, yielding a total of 439,163 reviews across $399$ unique companies, including 378 S\&P 500 companies. The average rating across companies ranges from a minimum value of $1.62$ up to a maximum value of $5$ ($ \mu = 3.37, \sigma = 0.40)$. 

\section{Data Representativeness}

We analyzed a total of  358,527 reviews of companies with at least 1000 reviews and presence in at least 10 U.S. states.
The period analyzed was between 2008 to 2020. All 51 U.S. states are represented in our sample (Table~\ref{tabsup:reviews_states}), with California (57,512 reviews) and  Wyoming (171 reviews) accounting for the highest and the lowest number of reviews respectively. The reviews span across 11 industries that were classified according to the Global Industry Classification Standard (GICS). Companies in the Consumer Discretionary industry accumulated the highest number of reviews, while companies operating in the Materials industry had the lowest (Table~\ref{tabsup:industries}). The reviews were written by managers, sales associates, software engineers, analysts, among others (Table~\ref{tabsup:employees_titles}).


\begin{table}[!htb]
\caption{Number of reviews and number of offices listed on the company reviewing site across U.S. States, ranked by the number of reviews published between 2008 and 2020 in descending order. Companies in the state of California accumulated the most published reviews, while companies based in Wyoming had the least published reviews. The Pearson correlation between the log number of reviews and the number of companies per state in our data is $.98$, while the correlation between the log number of reviews in our data and the log of population size across states is $.92$.}
\label{tabsup:reviews_states}
\begin{minipage}{0.5\textwidth}
\centering
\begin{tabular}{lrr}
\toprule
States &  \# Reviews &  \# Offices \\
\midrule
CA    &      57512 &        104 \\
TX    &      35208 &        104 \\
NY    &      31645 &        103 \\
IL    &      20313 &        103 \\
FL    &      20097 &        104 \\
GA    &      14437 &        103 \\
WA    &      14015 &        100 \\
NC    &      11812 &        102 \\
PA    &      11548 &        101 \\
AZ    &      10459 &        102 \\
MA    &       9839 &         99 \\
NJ    &       9574 &         99 \\
OH    &       9567 &        102 \\
VA    &       9480 &        101 \\
CO    &       7988 &        102 \\
MN    &       6725 &         93 \\
TN    &       6274 &         99 \\
MI    &       5982 &        100 \\
OR    &       5892 &         99 \\
MO    &       5601 &         99 \\
MD    &       5417 &         98 \\
IN    &       4262 &         98 \\
WI    &       3967 &         89 \\
KY    &       3788 &         94 \\
CT    &       3659 &         91 \\
DC    &       3222 &         93 \\
SC    &       2807 &         94 \\
KS    &       2776 &         92 \\
UT    &       2759 &         96 \\
OK    &       2426 &         88 \\
AL    &       2230 &         89 \\
LA    &       2042 &         91 \\
NV    &       1963 &         89 \\
NH    &       1427 &         84 \\
DE    &       1359 &         67 \\
IA    &       1358 &         82 \\
RI    &       1311 &         66 \\
AR    &       1263 &         82 \\
NE    &        993 &         87 \\
NM    &        944 &         79 \\
MS    &        934 &         73 \\
ID    &        676 &         69 \\
WV    &        568 &         68 \\
HI    &        491 &         56 \\
ME    &        382 &         71 \\
ND    &        290 &         53 \\
MT    &        285 &         50 \\
VT    &        278 &         44 \\
SD    &        267 &         45 \\
AK    &        244 &         42 \\
WY    &        171 &         43 \\
\bottomrule
\end{tabular}
\end{minipage}
\end{table}

\begin{figure*}
    \centering
    \includegraphics[width=0.55\linewidth]{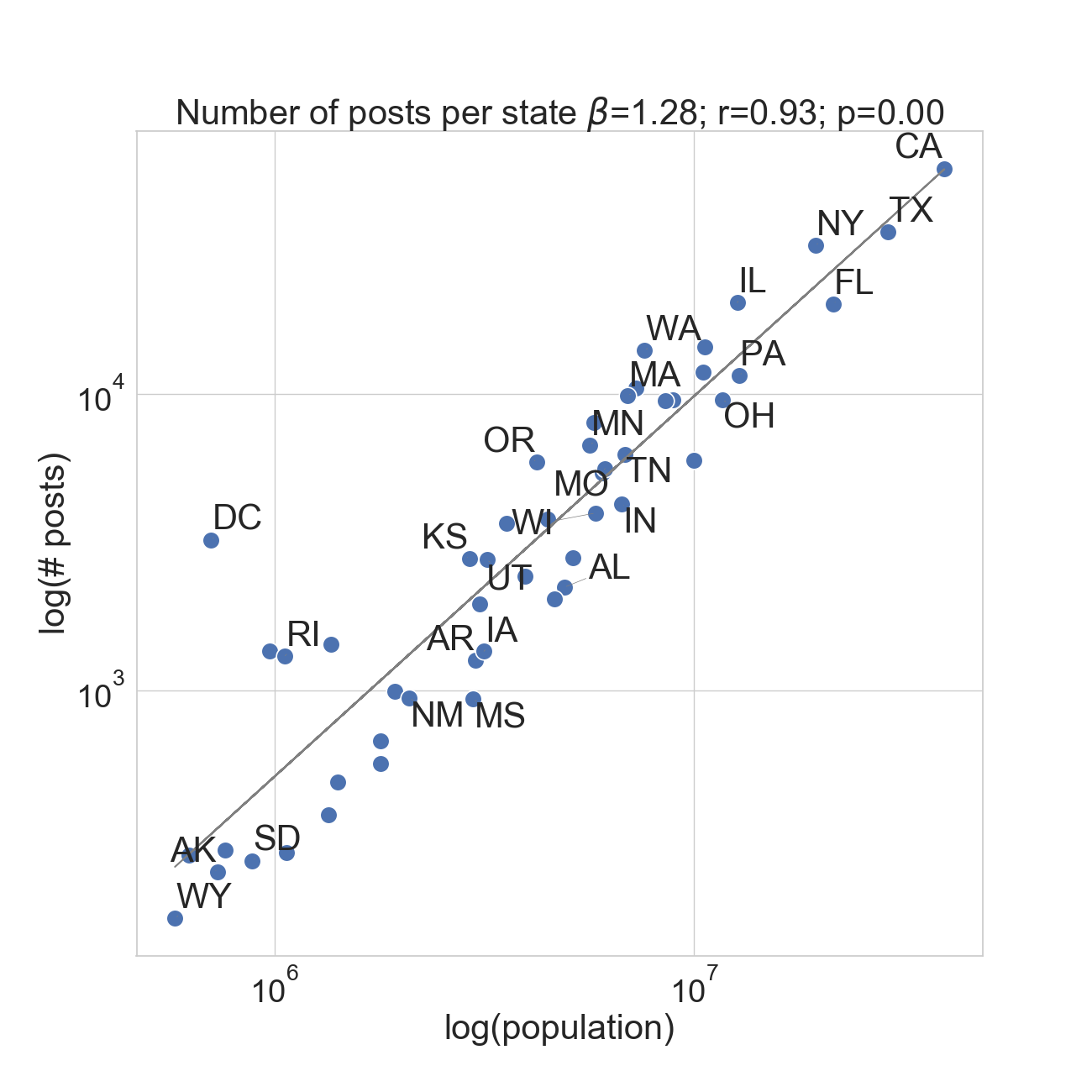}
    \caption{Number of reviews (log) in our dataset versus state population (log). The  states of Washington DC (DC) and Rhode Island (RI) have more reviews that what the population size would suggest. The line of best linear fit is shown in gray. U.S. states are shown with the two-code state abbreviation.}
    \label{fig:supp-spatial_rep}
\end{figure*}

\begin{table}[t!]
\caption{Number of reviews across different industries classified according to the Global Industry Classification Standard (GICS).}
\label{tabsup:industries}
\centering
\begin{tabular}{lrr}
\toprule
GICSSector &  \# Reviews &  \# Companies \\
\midrule
Consumer Discretionary &      55543 &           13 \\
Information Technology &      48677 &           12 \\
Financials             &      40116 &           11 \\
Health Care            &      26524 &           11 \\
Consumer Staples       &      21703 &            5 \\
Industrials            &      15568 &            7 \\
Communication Services &      10677 &            3 \\
Energy                 &       1468 &            1 \\
Materials              &       1045 &            1 \\
\bottomrule
\end{tabular}
\end{table}

\begin{table}[t!]
\caption{Number of reviews across roles (top 15) and employee status (top 5).}
\label{tabsup:employees_titles}
\centering
\begin{tabular}{lr}
\toprule
Employee Title &     \# Reviews \\
\midrule
 Sales Associate                 &  7786 \\
 Cashier                         &  3610 \\
 Manager                         &  3514 \\
 Software Engineer               &  3409 \\
 Customer Service Representative &  3202 \\
 Director                        &  1999 \\
 Store Manager                   &  1903 \\
 Assistant Manager               &  1837 \\
 Project Manager                 &  1802 \\
 Senior Manager                  &  1797 \\
 Associate                       &  1745 \\
 Pharmacy Technician             &  1739 \\
 Delivery Driver                 &  1584 \\
 Senior Software Engineer        &  1575 \\
 Sales Associate/Cashier         &  1498 \\
\toprule \textbf{Employee Status} &  \textbf{\# Reviews} \\
\midrule
Current Employee   &  143384 \\
Former Employee    &  111551 \\
Former Intern      &    5020 \\
Former Contractor  &    3183 \\
Current Intern     &    2789 \\
\bottomrule
\end{tabular}
\end{table}

\clearpage
\section{Methodological Details of our Approach for Detecting ISEs}

\begin{figure}[ht]
\centering
\includegraphics[width=\linewidth]{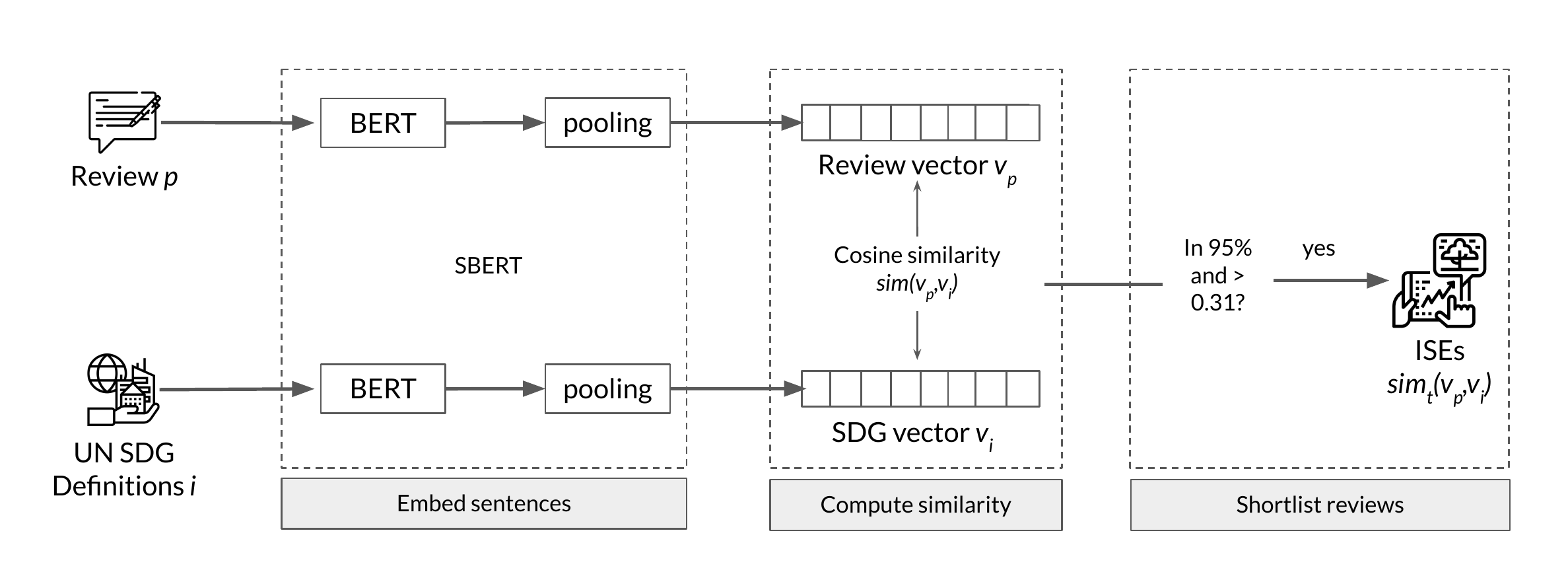}
\caption{\textbf{The deep learning framework detecting Internal Sustainability Efforts (ISE) from reviews.}}
\label{fig:ise_sbert}
\end{figure}

In what follows, we provide further details on the mixed-method approach employed to conceptualize and operationalize the construct of Internal Sustainability Efforts (ISEs) from the broader sustainability agenda embedded in the United Nations (UN) Sustainable Development Goals (SDGs) (Table~\ref{tab:defs}).

\subsection*{Step 1 - Human-driven pre-selection of goals}\label{sec:selecting_sdgs}



The SDGs are a collection of 17 interlinked global goals designed to be a ´´blueprint to achieve a better and more sustainable future for all". The SDGs were set up in 2015 by the United Nations General Assembly and are intended to be achieved by the year 2030. They are included in a UN Resolution called the 2030 Agenda. 

Specifically, the 17 SDGs are: (1) No Poverty, (2) Zero Hunger, (3) Good Health and Well-being, (4) Quality Education, (5) Gender Equality, (6) Clean Water and Sanitation, (7) Affordable and Clean Energy, (8) Decent Work and Economic Growth, (9) Industry, Innovation and Infrastructure, (10) Reducing Inequality, (11) Sustainable Cities and Communities, (12) Responsible Consumption and Production, (13) Climate Action, (14) Life Below Water, (15) Life On Land, (16) Peace, Justice, and Strong Institutions, and (17) Partnerships for the Goals. 

Given their broad scope, not all 17 UN goals might be relevant to internal corporate practices. To decide whether to retain or discard some UN SDGs, three independent annotators qualitatively assessed the definition and scope of each goal. They unanimously decided to discard the following four: `life below water', `life on land', `sustainable cities', and `partnerships for goals'. More precisely, `life below water', `life on land', `sustainable cities' mainly focus on the health of water bodies, land conservation, and cities, respectively. With the exception of highly specialized companies focused on the conservation of water bodies, land conservation, or cities, these goals are unlikely to be featured in employees' reviews. `Partnership on goals', on the other hand, was explicitly designed to foster collaboration between countries to facilitate sustainability. Since we focused entirely on US-based companies, we also excluded any goals that pertain to international co-operations.
This initial qualitative stage resulted in the selection of 13 UN SDGs.

\subsection*{Step 2 - Unsupervised discovery of relevant reviews and goals}\label{sec:method}

\begin{table}[]
\centering
\small
\begin{tabular}{@{}
>{\columncolor[HTML]{FFFFFF}}l 
>{\columncolor[HTML]{FFFFFF}}l 
>{\columncolor[HTML]{FFFFFF}}l @{}}
\toprule
id    & Goal                                     & Definition                                                                                                                                                                                                                                \\ \midrule
\textcolor{gray}{SDG1}  & \textcolor{gray}{no poverty}                               & \textcolor{gray}{to end poverty in all its forms, everywhere}                                                                                                                                                                                               \\
\textcolor{gray}{SDG2}  & \textcolor{gray}{zero hunger}                              & \begin{tabular}[c]{@{}l@{}}\textcolor{gray}{End hunger, achieve food security and improved nutrition}\\  \textcolor{gray}{and promote sustainable agriculture}\end{tabular}                                                                                                   \\
SDG3  & \textbf{good health and wellbeing}                & To ensure healthy lives and promote well-being for all at all ages.                                                                                                                                                                       \\
SDG4  & \textbf{quality education}                        & \begin{tabular}[c]{@{}l@{}}Ensure inclusive and equitable quality education and promote\\ lifelong learning opportunities for all\end{tabular}                                                                                            \\
SDG5  & \textbf{gender equality}                          & Achieve gender equality and empower all women and girls                                                                                                                                                                                   \\
\textcolor{gray}{SDG6}  & \textcolor{gray}{clean water and sanitation}              & \begin{tabular}[c]{@{}l@{}}\textcolor{gray}{Ensure availability and sustainable management of water and}\\ \textcolor{gray}{sanitation for all.}\end{tabular}                                                                                                                 \\
\textcolor{gray}{SDG7}  & \textcolor{gray}{affordable and clean energy}              & \begin{tabular}[c]{@{}l@{}}\textcolor{gray}{Ensure access to affordable, reliable, sustainable and modern}\\ \textcolor{gray}{energy for all.}\end{tabular}                                                                                                                   \\
SDG8  & \textbf{decent work and economic growth}          & \begin{tabular}[c]{@{}l@{}}Foster sustained, inclusive and sustainable economic growth,\\  full and productive employment and decent work for all.\end{tabular}                                                                           \\
SDG9  & \textbf{industry, innovation, and infrastructure} & \begin{tabular}[c]{@{}l@{}}build resilient infrastructure, promote sustainable industrialization\\ and foster innovation\end{tabular}                                                                                                     \\
\textcolor{gray}{SDG10} & \textcolor{gray}{reducing inequality}                      & \textcolor{gray}{Reduce inequality within and among countries}                                                                                                                                                                                              \\
\textcolor{gray}{SDG11} & \textcolor{gray}{sustainable cities and communities}       & \textcolor{gray}{Make cities inclusive, safe, resilient and sustainable}                                                                                                                                                                                    \\
\textcolor{gray}{SDG12} & \textcolor{gray}{responsible consumption and production}   & \textcolor{gray}{To ensure sustainable consumption and production patterns}                                                                                                                                                                                 \\
\textcolor{gray}{SDG13} & \textcolor{gray}{climate action}                           & \textcolor{gray}{Take urgent action to combat climate change and its impacts}                                                                                                                                                                               \\
\textcolor{gray}{SDG14} & \textcolor{gray}{life below water}              & \begin{tabular}[c]{@{}l@{}}\textcolor{gray}{Conserve and sustainably use the oceans, seas and marine resources}\\ \textcolor{gray}{for sustainable development}\end{tabular}                                                                                                  \\
\textcolor{gray}{SDG15} & \textcolor{gray}{life on land}                             & \begin{tabular}[c]{@{}l@{}}\textcolor{gray}{Protect, restore and promote sustainable use of terrestrial}\\ \textcolor{gray}{ecosystems, sustainably manage forests, combat desertification,} \\ \textcolor{gray}{and halt and reverse land degradation and halt biodiversity loss}\end{tabular} \\
SDG16 & \textbf{peace, justice, and strong institutions}   & \begin{tabular}[c]{@{}l@{}}Promote peaceful and inclusive societies for sustainable\\ development, provide access to justice for all and build effective,\\ accountable and inclusive institutions at all levels\end{tabular}             \\
\textcolor{gray}{SDG17} & \textcolor{gray}{partnerships for the goals}               & \begin{tabular}[c]{@{}l@{}}\textcolor{gray}{Strengthen the means of implementation and revitalize the global}\\ \textcolor{gray}{partnership for sustainable development}\end{tabular}                                                                                        \\ \bottomrule
\end{tabular}
\caption{The 17 UN Sustainability Development Goals (SDGs) and their definitions as put forth by the UN~\cite{wced1987world}. The six goals in bold (SDGs 3,4,5,8,9, and 16) are our six ISEs, while the remaining ones did not apply to the internal corporate context and, as such, are  in gray.}
\label{tab:defs}
\end{table}

We obtained mentions of similarity in reviews by seeing how semantically related the review sentences were to the definitions of the remaining 13 SDG goals. To find the similarity between reviews and goal definitions, we employed a deep-learning method that is tailored to find the similarity between two sentences called sentence BERT or SBERT~\cite{reimers2019sentence}, which is summarized in Figure~\ref{fig:ise_sbert}. The Bidirectional Encoder Representations from Transformers (BERT)~\cite{devlin2018bert}, or its variants like RoBERTa and DistilBERT, is a family of state-of-the-art Natural Language Processing (NLP) methods that are trained on a vast corpora of data, enabling them to learn many different types of language phenomena. One such phenomenon is semantic similarity between different sentences. SBERT is trained especially for this task and has achieved state-of-the-art results in text similarity NLP tasks~\cite{reimers2019sentence}.

\begin{figure*}[!t]
    \centering
    \includegraphics[width=\linewidth]{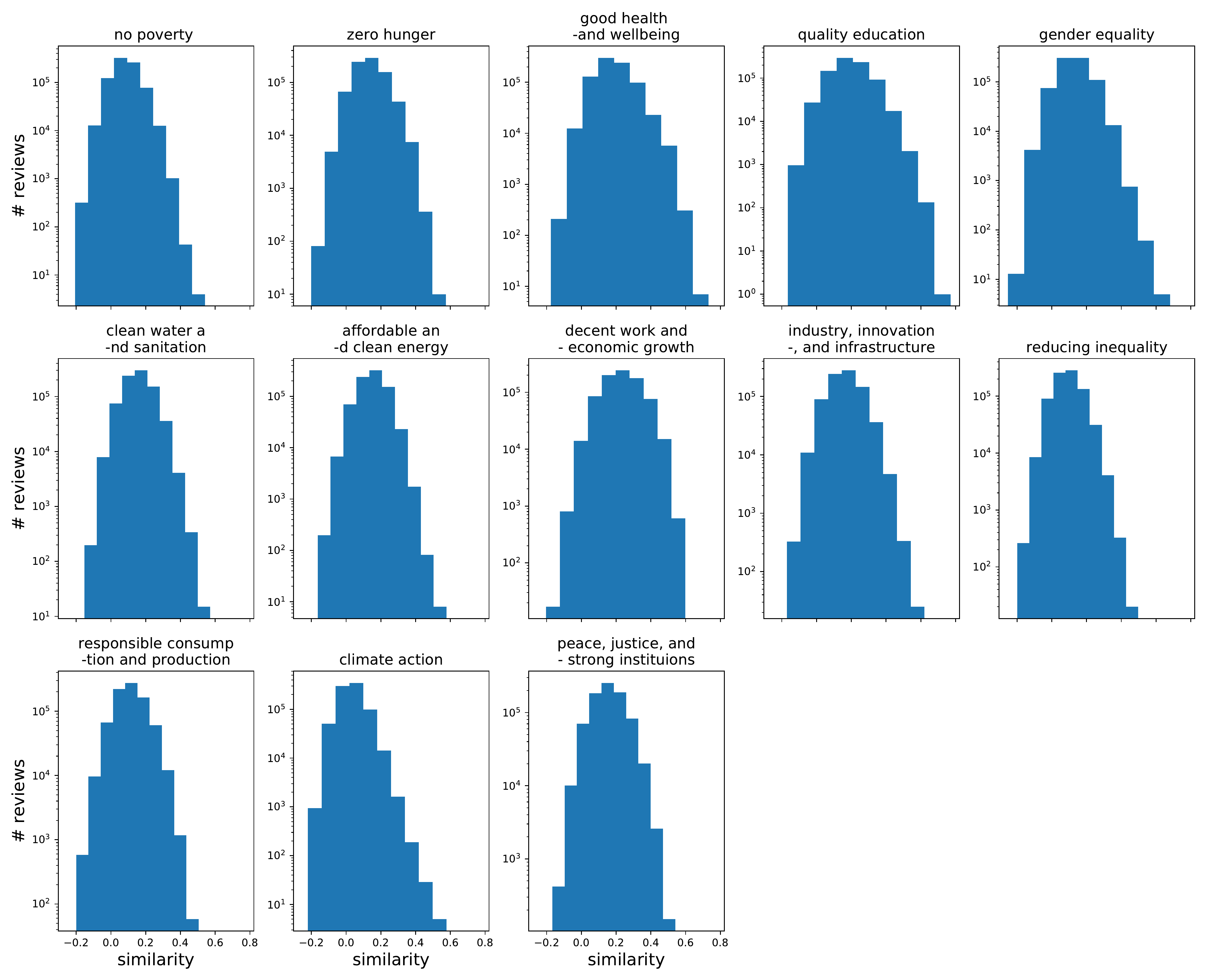}
    \caption{The distribtion of similarity scores for all 13 SDGs shortlisted after the conceptual check (Step 1).}
    \label{fig:sim_dist}
\end{figure*}

We embedded each of the 13 UN SDG definitions (Table~\ref{tab:defs}) using SBERT to obtain 13 vectors of length 726 ($v_i$ for the $i^{th}$ SDG). 
Each review $p$ consists of a title, pros, and cons. The pros and cons can be several sentences long, and, as we wanted to precisely capture the presence of ISEs, we split each review into individual sentences $s^1$, $s^2$,...$s^k$ using a sentence tokenizer. We embedded each of these sentences as well with SBERT ($v_j$ for the $j^{th}$ sentence in $p$). We then obtained the cosine similarity between $v_j$ and $v_i$. Finally, since we required sustainability labels and scores at \textit{review} level, we aggregated the sentence-level similarity scores by denoting the score of a review to be its highest scoring sentence:

\begin{equation}
   sim(v_p,v_i)= cosine[v_j,v_i] \textrm{ such that } 
   sim(v_j,v_i) \textrm{ is maximum } \forall \textrm{ } j^{th} \textrm{ sentence in } p
\end{equation}

\noindent where $p$ is each post, $i$ is the definition of the $i^{th}$ UN sustainability goal, $v_p$ is the SBERT vector embedding of post $p$, $v_i$ is the SBERT vector embedding  of $i$, and $v_j$ is the SBERT vector embedding of the $j^{th}$ sentence in post $p$. We ended up with similarity scores ranging from -1 and 1 for all sentences for all 13 goals. The distribution of the scores is summarized in Figure~\ref{fig:sim_dist}. The similarity distributions are different for different goals; e.g., `decent work' has higher average similarity compared to `gender equality'. 

We considered whether to use pros, cons, or both to understand and operationalize the ISE construct. As sustainability is positively valenced, we hypothesized that company-led efforts on sustainability would be appreciated or brought up more frequently in pros rather than cons. Furthermore, our analysis revealed that the average similarity for pros was much higher than the average similarity for cons (Table~\ref{tab:pros_vs_cons}), thus indicating that our NLP method is more effective when assessing sustainability concerns for pros compared to cons.
This was also confirmed through a qualitative analysis involving 3 independent annotators who assessed the top most similar cons and found them to be not relevant to the ISE they were picked for (Fleiss $k$ = 0.91).

\begin{table}
\centering
\begin{tabular}{@{}lrrrr@{}}
\toprule
                          & \multicolumn{2}{l}{\begin{tabular}[c]{@{}l@{}}avg SBERT \\ similarity\end{tabular}} & \multicolumn{2}{l}{\begin{tabular}[c]{@{}l@{}}proportion of \\ relevant reviews\end{tabular}} \\ \midrule
Goal                      & pros                                      & cons                                     & pros                                          & cons                                          \\ \midrule
Monetary        & \textbf{0.234}                                     & 0.153                                    & 0.180                                         & \textbf{0.202}                                         \\
Health                    & \textbf{0.185}                                     & 0.110                                    & \textbf{0.177}                                         & 0.084                                         \\
Education                 & \textbf{0.194}                                     & 0.114                                    & \textbf{0.175}                                         & 0.112                                         \\
Diversity           & \textbf{0.130}                    & 0.080                                    & \textbf{0.130}                                         & 0.026                                         \\
Infrastructure & \textbf{0.163}                                     & 0.113                                    & \textbf{0.174}                                         & 0.084                                         \\
Atmosphere    & \textbf{0.157}                                     & 0.102                                    & \textbf{0.175}                                         & 0.058                                         \\ \bottomrule
\end{tabular}
\caption{Average similarity scores for pros and cons based on SBERT similarity, and the proportion of reviews shortlisted based on our deep learning method for pros and cons.}
\label{tab:pros_vs_cons}
\end{table}


To then understand whether the above similarity metric correctly captured mentions of the pre-selected 13 SDGs in the corporate context, three independent annotators manually assessed the five highest ranked sentences for each goal based on their similarity score for that goal. By assessing the top five, we reached an understanding of the upper bound of our method. Agreement between annotators, measured using Fleiss Kappa, was high (0.83). For disagreements, we used the majority rating to obtain a final ground-truth label for relevance. We retained only those goals for which at least four of the top five highest ranked sentences were relevant to the goal (i.e, they mentioned concepts related to that goal, such as, mentions of gender diversity initiatives in the company). We found that the goals related to environmental sustainability, `clean water' and `climate action', faced word sense disambiguation issues with shortlisted sentences describing aspects of the \textit{work environment} that are not pertinent to scope of this analysis (e.g., the cleanliness of office spaces). 
Thus, we discarded a further five UN goals in this step (clean energy, clean water, climate action, responsible consumption, and no poverty) and retained a total of eight goals. 

To ensure high precision while accounting for the fact that the the similarity $sim(v_p, v_i)$ frequency distributions are different for different goals, we opted for a two-step selection procedure. To shortlist reviews, we followed an approach based on previous literature~\cite{das2020modeling}, and
to filter out reviews that do not mention an ISE, we again used the similarity score. We formulated an SBERT similarity score, $sim_t(v_p,v_i)$, that measures whether and the extent to which a post $p$ is about ISE $i$ with the following function:

\begin{equation}
    sim_t(v_p,v_i) = \begin{cases} sim(v_p,v_i), & \text{if } sim(v_p,v_i)>0.31 \textrm{ AND } sim(v_p,v_i)>95\%(i)
    \\
    0,              & \text{otherwise}
    \end{cases}
\end{equation}

where $p$ is the post, $v_p$ is the vector embedding the post, and  $v_i$ is the vector embedding the UN's definition of practices around $i$. The post is about $i$ if two conditions are met: the post vector, $v_p$'s similarity to $v_i$ is above a fixed threshold 0.31, and greater than the 95\% similarity value. The 0.31 threshold is the average of $sim(v_p, v_i)$ at the 95\% threshold for the 8 goals selected at this stage.  The last two columns of table~\ref{tab:pros_vs_cons} show the proportion of selected pros and cons based on $sim_t(v_p, v_i)$. Note that the proportion of shortlisted reviews for cons were much lower than those shortlisted by pros (Table~\ref{tab:pros_vs_cons}).  Therefore, we further confirmed that pros are more appropriate for understanding ISEs conceptually and empirically. For the rest of the analysis, we utilize pros only.

\subsection*{Step 3 - Consolidation of goals}

\begin{figure}[!ht]
\centering
\includegraphics[width=\linewidth]{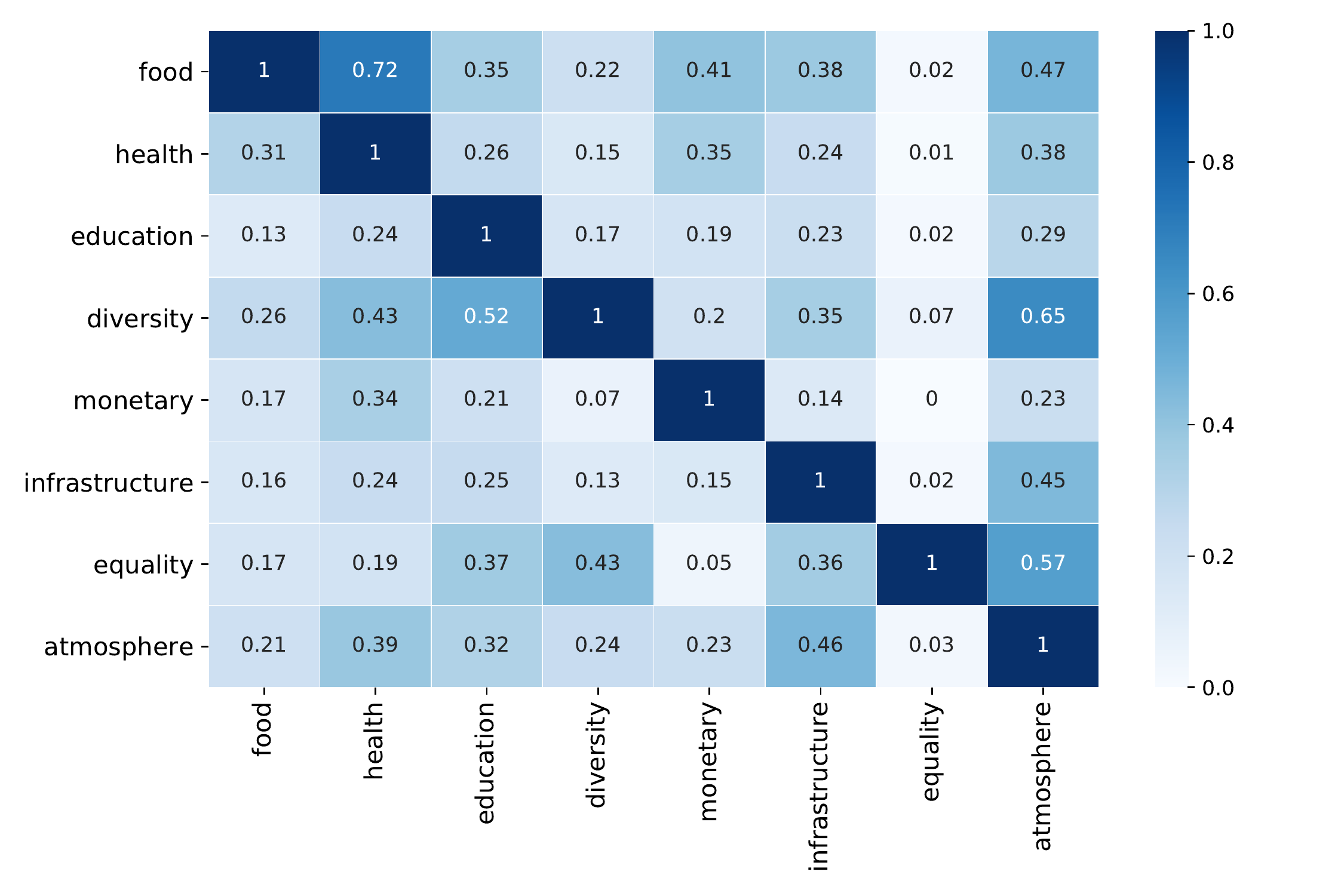}
\caption{Overlap between goal pair. Step 3 of the goal selection process checked for content overlap between each pair of goals.}
\label{fig:goal_overlap}
\end{figure}

Sustainability goals are not mutually exclusive and a certain degree of overlap might be expected (e.g., 
work-life balance facilitates both health and gender equality, and is therefore a concept shared by both ISEs). However, there might be cases where two goals are so strongly related to one another that cannot be discerned from each other. To systematically tackle the issue of semantically overlapping goals, we plotted the content overlap $O$ for each pair of goals by computing the proportion of sentences that the two goals $j$ and $k$ have in common (Figure~\ref{fig:goal_overlap}):

\begin{equation}
    O(j, k) = \frac{|R(j, u) \cap R(k, u)|}{|R(j, u)|}
\end{equation}

\noindent where $R(j, u) = [p \in R(u)$, if $sim_t(v_p,v_j)>0]$, which is the set of $u$'s reviews relevant to goal $j$; and $R(k, u) = [p \in R(u)$, if $sim_t(v_p,v_k)>0]$,  which is the set of $u$'s reviews relevant to goal $k$. The ordering of the goals in the overlap function $O()$ impacts the denominator (the first goal goes to the denominator), and that is why $O(j, k)$ is  a non-symmetric metric.




We observe that the only overlap higher than $0.5$ occurs for the pair `food (no hunger)' \emph{vs} `health'. These have indeed strong conceptual relatedness in the corporate sector, and thus we proceeded by subsuming `no hunger' under `health'. 


We note that two other pairs of goals exhibited semantic relatedness close to 0.5: these were `supportive environment' \emph{vs} `supporting infrastructure', and `diversity' vs `gender equality'. To decide whether to combine or keep these pairs separate, the three annotators qualitatively assessed the top five reviews for each goal. Annotators found `supportive environment' and `supporting infrastructure' to cover related yet different concerns; however, they discovered that the `diversity' goal (reducing inequality) was mostly expressed through mentions of `gender discrimination', thus becoming almost indistinguishable from the concerns raised for the other goal `gender equality'. Consequently, we merged these two goals together to account for the identified conceptual overlaps. 

In addition to Equation (1) in the main manuscript, we  tested two other variants of the linear score $s(u, i)$, one exponentially increasing with similarity $sim_t$ and the other logarithmically, to score  $u$ (company or state) in terms of the $i^{th}$ ISE:

\begin{equation*}
    s(u, i) = \frac{\sum_{p \in R(u)} \frac{e^{sim_t(v_p,v_i)}}{e}}{|R(u)|} 
\end{equation*}

or

\begin{equation*}
    s(u, i) = \frac{\sum_{p \in R(u)} \frac{log{(sim_t(v_p,v_i)+1)}}{log(2)}}{|R(u)|} 
\end{equation*}

where $R(u)$ is the number of  reviews at study unit $u$. We found that the two variations had results  similar to the linear scaling in our linguistic validation. As such, in the main manuscript, we reported the results for the simplest, linear scoring. 


\section{Method Validation}

In the following section we provide further evidence of the process followed to validate our deep learning method for detecting ISEs. 

\noindent
\subsection*{ISEs and Online Ratings. } On the company review platform, employees have the option to rate the companies they are reviewing on five different facets --- culture, balance, company, management, career, and an overall score. 
After aggregating the ISE scores at a company level, we found statistically significant positive correlation between our six ISE scores and the company's ratings (Table \ref{tab:glassdoor_rating_weighted}). To see whether the correlations were merely an artifact of company popularity, we also computed the correlation coefficient between the total number of reviews of a company, our six ISEs scores, and the six company ratings (Table \ref{tab:glassdoor_rating_weighted}).

The number of company reviews were not correlated with the ratings, while being slightly (but usually not significantly) correlated with our scores. This indicates that our correlations are indeed capturing the relationship between our scores and the online ratings, rather than capturing overall company popularity. Our health ISE is most strongly correlated with balance $[r = 0.70, p < 0.001]$. Since the health ISE captures mentions of work-life balance, this finding supported our conceptualization. The education, diversity, infrastructure, and atmosphere ISEs were strongly and significantly associated with the career rating. Education and training opportunities as well as infrastructure facilitate career growth~\cite{morgan2017employee}. Monetary was strongly correlated with the overall company rating $[r = 0.75, p < 0.001]$, in line with previous research that found the importance of salary in company evaluations~\cite{cascio2006decency}. Finally, atmosphere is strongly associated with culture $[r = 0.71, p < 0.001]$ and management $[r = 0.66, p < 0.001]$, thus indicating that this ISE captures relevant dimensions of corporate life.



\begin{table}[!t] 
\centering
\small
\begin{tabular}{llllllll}
\hline
Rating                                                            & Monetary         & Health           & Education        & Diversity        & Infrastructure   & Atmosphere       & \begin{tabular}[c]{@{}l@{}}Total Reviews \\ (logged)\end{tabular} \\ \hline
Culture                                                           & 0.29***          & 0.52***          & 0.57***          & 0.63***          & 0.66***          & 0.71***          & -0.09                                                             \\
Balance                                                           & 0.48***          & \textbf{0.70***} & 0.32***          & 0.56***          & 0.58***          & 0.65***          & -0.15                                                             \\
Management                                                        & 0.26**           & 0.43***          & 0.57***          & 0.62***          & 0.64***          & 0.66***          & -0.06                                                             \\
Career                                                            & 0.40***          & 0.49***          & \textbf{0.70***} & \textbf{0.66***} & \textbf{0.70***} & \textbf{0.72***} & -0.08                                                             \\
Overall                                                           & \textbf{0.75***} & 0.66***          & 0.55***          & 0.61***          & 0.64***          & \textbf{0.72***} & -0.11                                                             \\ \hline
\begin{tabular}[c]{@{}l@{}}Total Reviews \\ (logged)\end{tabular} & -0.15            & -0.29***         & -0.03            & -0.22*           & -0.18            & -0.24*           & 1.00***                                                           \\ \hline
\end{tabular}
\caption{
\textbf{Pearson correlation between each company's five online reviews and its ISE scores.} The highest correlation with a rating for each ISE is marked in bold. (*** for p $<$ 0.005, ** for p $<$ 0.01, * for p $<$ 0.05) 
}
\label{tab:glassdoor_rating_weighted}
\end{table}

\begin{table}[]
\small
\centering
\begin{tabular}{@{}lllll@{}}
\toprule
                  & Gender report & \begin{tabular}[c]{@{}l@{}}Fashion report \\ (Gender Diversity)\end{tabular} & \begin{tabular}[c]{@{}l@{}}Fashion report \\ (Financial Benefit)\end{tabular} & \begin{tabular}[c]{@{}l@{}}Fashion report \\ (Supportive \\ Environment)\end{tabular} \\ \midrule
$s(u,i)$          & 0.285         & 0.285                                                                             &   0.236                                                                            & 0.246                                                                                      \\
random (baseline) & 0.186         & 0.183                                                                              & 0.185                                                                               & 0.185                                                                                      \\ \bottomrule
\end{tabular}
\caption{Ranked Biased Overlap (RBO) scores measuring the concordance between review-based sustainability rankings by $s(u,i)$ and four external rankings. Higher RBO scores are better. As a baseline, we compare our rankings against a random ordering of companies (averaged over 1000 runs).}
\label{tab:external_reports}
\end{table}

\begin{table}[]
\small
\centering
\begin{tabular}{@{}cccccc@{}}
\toprule
                  &               & \multicolumn{4}{c}{Fashion Report}                                                                                \\ \midrule
                  & Gender report & Gender Diversity & Financial Benefit & \begin{tabular}[c]{@{}c@{}}Supportive \\ Environment\end{tabular} & Health \\ \midrule
$s(u,i)$          & 0.22          & 0.073            & 0.100               & 0.227                                                             & 0.136  \\
random (baseline) & 0.001         & -0.012           & 0.001            & -0.015                                                            & 0.002  \\ \bottomrule
\end{tabular}
\caption{Spearman rank correlation measuring the overlap between review-based sustainability rankings by $s(u,i)$ and five external rankings. As a baseline, we compare our rankings against a random ordering of companies (averaged over 1000 runs).}
\label{tab:external_reports_spearmanr}
\end{table}

\mbox{ }\\
\noindent
\subsection*{ISEs and external reports.} We used external sustainability reports to further assess the validity of our method. Reports on sustainability are few and fragmented~\cite{burritt2010sustainability}, making it challenging to establish validity through external sources. However, we obtained two external reports on sustainability and compared the ranking of companies in them with our review-based ISE rankings. Overall, results of these comparisons found that our method substantially outperformed a random baseline. Specifically, we utilized two reports. The first report contained gender diversity indicators of S\&P 500 companies~\cite{equileap}, providing a ranking of 25 companies performing strongly in terms of gender equality ($gr(u)$ is the score of company $u$ in the gender report). The second report contained the Fashion Transparency Index~\cite{fti}, which scores companies in the fashion sector on different sustainability metrics such as `Discrimination' and `Diversity and Inclusion' (called the \emph{fashion report} from now on). Three annotators mapped those metrics into our six ISEs (Table~\ref{tab:fti_mapping}). We then computed the score of company $u$ for the $i^{th}$ ISE in the fashion report ($fr(u,i)$) by averaging $u$'s scores for the metrics mapped to the $i^{th}$ ISE.  With those two scores at hand, we then tested whether our six ISEs captured the constructs they were meant to capture. We ranked companies by their $gr(u)$, ranked them by their $s(u,i=`gender')$, and computed the correlations between these two lists. To then go beyond gender, for each $i^{th}$ ISE,  we ranked companies by their $fr(u,i)$, ranked them by their $s(u,i)$, and computed the correlations between these two lists. Since the ranked lists might have included different companies, we utilized a measure called `rank biased overlap' (RBO), which was designed for measuring the goodness of ranking between non-overlapping ranked lists~\cite{webber2010similarity} (Table \ref{tab:external_reports}). RBO scores lie between 0 and 1, with a higher score representing better concordance between lists. We also calculated the Spearman correlation between the ranking of the companies that are common in both lists (Table \ref{tab:external_reports_spearmanr}). Overall, ranking companies by $s(u, i)$ outperformed the random baseline not only for the gender ISE but also for the remaining five ISEs.


\section{Distribution of Similarity Score}

\clearpage

\section{Stock Analysis}

The geometric mean of stock growth is calculated as $GM({\textrm{stock growth}_{[09-19]}}) = \Pi (\textrm{stock growth}_{[09-19]}(c))^{1/n} $, where $c$ is a company in a specific \emph{(ISE facet (staff welfare or financial benefits), percentile)} bin, and $n$ is the number of the companies in such a bin.

\begin{figure*}[!t]
    \centering
    \includegraphics[width=\linewidth]{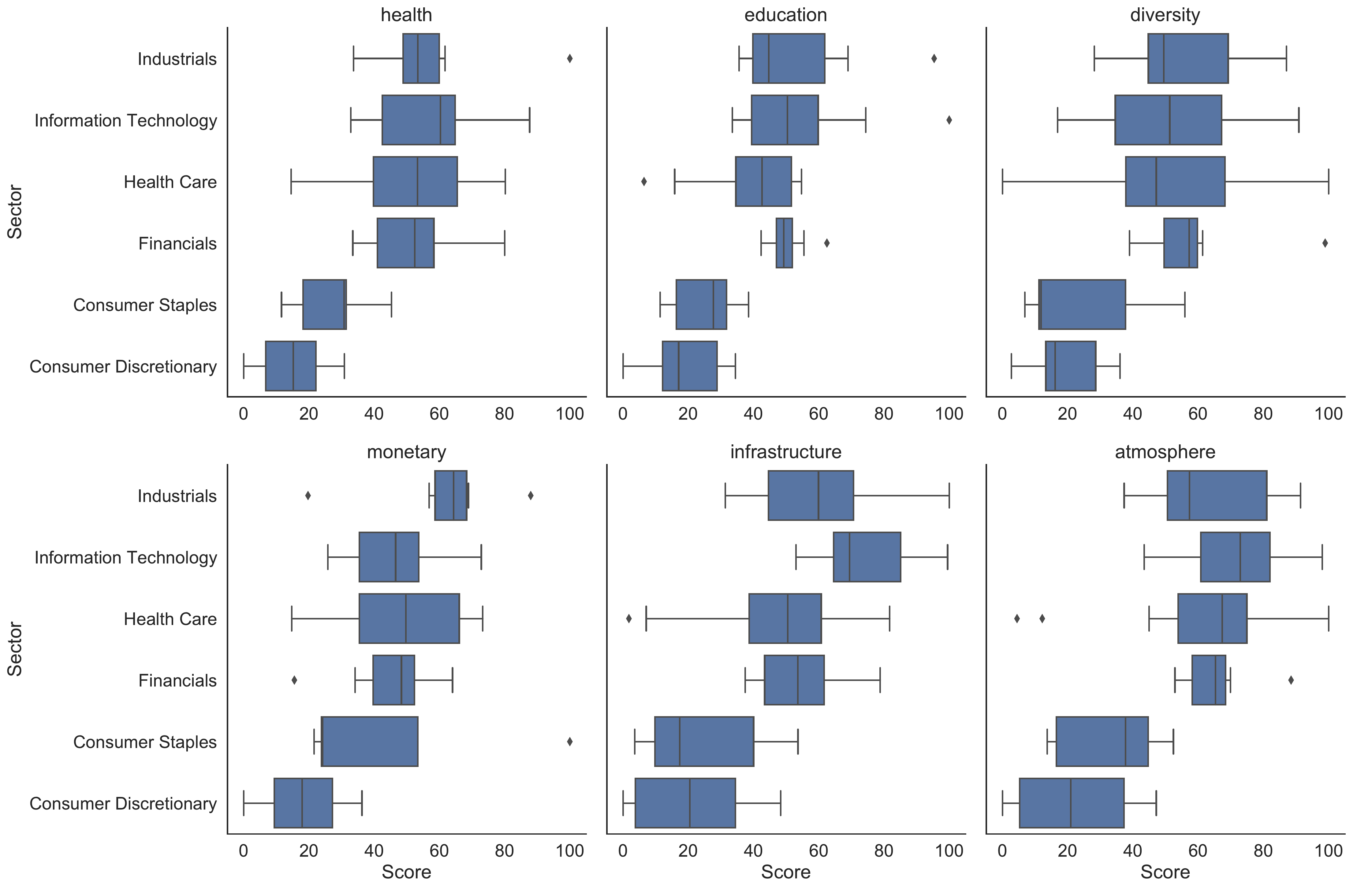}
    \caption{Scores for the six ISEs across industry sectors.}
    \label{fig:sector_all}
\end{figure*}

\begin{table}[]
\centering
\scriptsize
\begin{tabular}{@{}lllllll@{}}
\toprule
                                                                                                                            & Monetary & Health & Education & Diversity & Infrastructure & Atmosphere \\ \midrule
Animal Welfare                                                                                                              &          &        &           &           &                &            \\
Annual Leave \& Public Holidays                                                                                             & x        &        &           &           &                & x          \\
Anti-bribery, Corruption, \& Presentation of False Information                                                              &          &        &           &           &                &            \\
Biodiversity \& Conservation                                                                                                &          &        &           &           &                &            \\
Community Engagement                                                                                                        &          &        &           &           &                & x          \\
Contracts \& Terms of Employment                                                                                            & x        &        &           &           &                &            \\
Discrimination                                                                                                              &          &        &           & x         &                &            \\
Diversity \& Inclusion                                                                                                      &          &        &           & x         &                &            \\
Energy \& Carbon Emissions                                                                                                  &          &        &           &           &                &            \\
Equal Pay                                                                                                                   & x        &        &           & x         &                &            \\
Freedom of Association, Right to Organise \& Collective Bargaining                                                          &          &        &           &           &                & x          \\
Harassment \& Violence                                                                                                      &          &        &           & x         &                &            \\
Health \& Safety                                                                                                            &          & x      &           &           &                &            \\
Maternity Rights \& Parental Leave                                                                                          &          &        &           & x         &                &            \\
Notice Period, Dismissal \& Disciplinary Action                                                                             &          &        &           &           &                & x          \\
Restricted Substances List                                                                                                  &          &        &           &           &                &            \\
\begin{tabular}[c]{@{}l@{}}Wages \& Financial Benefits (e.g. bonuses, insurance,\\  social security, pensions)\end{tabular} & x        &        &           &           &                &            \\
Waste \& Recycling (Packaging/Office/Retail)                                                                                &          &        &           &           &                &            \\
Waste \& Recycling (Product/Textiles)                                                                                       &          &        &           &           &                &            \\
Water Usage \& Footprint                                                                                                    &          &        &           &           &                &            \\
Working Hours \& Rest Breaks                                                                                                & x        & x      &           & x         &                & x          \\ \bottomrule
\end{tabular}
\caption{Mapping between the fashion report's metrics~\cite{fti} and our 6 ISEs. Note that some metrics are applicable for multiple goals.}
\label{tab:fti_mapping}
\end{table}

\end{document}